\documentclass[review]{elsarticle}

\usepackage{lineno}
\usepackage{url}
\usepackage[tight,footnotesize]{subfigure}
\modulolinenumbers[5]

\usepackage{multirow}
\usepackage{longtable}
\usepackage{booktabs}

\journal{Elsevier Journal}

\biboptions{authoryear}

\begin{document}

\begin{frontmatter}

\title{Applications of deep learning in stock market prediction: recent progress}

\author{Weiwei Jiang\corref{mycorrespondingauthor}}
\address{Department of Electronic Engineering, Tsinghua University, Beijing 100084, China}
\cortext[mycorrespondingauthor]{Corresponding author. E-mail address: jiangweiwei@mail.tsinghua.edu.cn}

\begin{abstract}
Stock market prediction has been a classical yet challenging problem, with the attention from both economists and computer scientists. With the purpose of building an effective prediction model, both linear and machine learning tools have been explored for the past couple of decades. Lately, deep learning models have been introduced as new frontiers for this topic and the rapid development is too fast to catch up. Hence, our motivation for this survey is to give a latest review of recent works on deep learning models for stock market prediction. We not only category the different data sources, various neural network structures, and common used evaluation metrics, but also the implementation and reproducibility. Our goal is to help the interested researchers to synchronize with the latest progress and also help them to easily reproduce the previous studies as baselines. Base on the summary, we also highlight some future research directions in this topic.
\end{abstract}


\end{frontmatter}


\section{Introduction}
    \label{sec:intro}
Stock market prediction is a classical problem in the intersection of finance and computer science. For this problem, the famous efficient market hypothesis (EMH) gives a pessimistic view and implies that financial market is efficient~\citep{fama1965behavior}, which maintains that technical analysis or fundamental analysis (or any analysis) would not yield any consistent over-average profit to investors. However, many researchers disagree with EMH ~\citep{malkiel2003efficient}. Some studies are trying to measure the different efficiency levels for mature and emerging markets, while other studies are trying to build effective prediction models for stock markets, which is also the scope of this survey.

The effort starts with the stories of fundamental analysis and technical analysis. Fundamental analysis evaluates the stock price based on its intrinsic value, \textit{i.e.}, fair value, while technical analysis only relies on the basis of charts and trends. The technical indicators from experience can be further used as hand-crafted input features for machine learning and deep learning models. Afterwards, linear models are introduced as the solutions for stock market prediction, which include autoregressive integrated moving average (ARIMA)~\citep{hyndman2018forecasting} and generalized autoregressive conditional heteroskedasticity (GARCH)~\citep{bollerslev1986generalized}. With the development of machine learning models, they are also applied for stock market prediction, \textit{e.g.}, Logistic regression and support vector machine~\citep{alpaydin2014introduction}.

Our focus in this survey would be the latest emerging deep learning, which is represents by various structures of deep neural networks~\citep{goodfellow2016deep}. Powered by the collection of big data from the Web, the parallel processing ability of graphics processing units (GPUs), and the new convolutional neural network family, deep learning has achieved a tremendous success in the past few years, for many different applications including image classification~\citep{rawat2017deep, jiang2020edge}, object detection~\citep{zhao2019object}, time series prediction~\citep{brownlee2018deep, jiang2018geospatial}, etc. With a strong ability of dealing with big data and learning the nonlinear relationship between input features and prediction target, deep learning models have shown a better performance than both linear and machine learning models on the tasks that include stock market prediction.

In the past few years, both the basic tools for deep learning and the new prediction models are undergoing a rapid development. With the continuous improved programming packages, it becomes easier to implement and test a novel deep learning model. Also, the collection of online news or twitter data provides new sources of predicting stock market. More recently, graph neural networks using various knowledge graph data appear as new ideas. The study for stock market prediction is not limited to the academia. Attracted by the potential profit by stock trading powered by the latest deep learning models, asset management companies and investment banks are also increasing their research grant for artificial intelligence which is represented by deep learning models nowadays.

Since there are many new developments in this area, this situation makes it difficult for a novice to catch up with the latest progress. To alleviate this problem, we summarize the latest progress of deep learning techniques for stock market prediction, especially those which only appear in the past three years. We also present the trend of each step in the prediction workflow in these three years, which would help the new-comers to keep on the right track, without wasting time on obsolete technologies.

We focus on the application of stock market, however, machine learning and deep learning methods have been applied in many financial problems. It would be beyond the scope of this survey to cover all these problems. However, the findings presented in this survey would also be insightful for other time series prediction problems in the finance area, \textit{e.g.}, exchange rate or cryptocurrency price prediction.

We also pay a special attention to the implementation and reproducibility of previous studies, which is often neglected in similar surveys. The list of open data and code from published papers would not only help the readers to check the validity of their findings, but also implement these models as baselines and make a fair comparison on the same datasets. Based on our summary of the surveyed papers, we try to point out some future research directions in this survey, which would help the readers to choose their next movement. 

Our main contribution in this survey are summarized as follows:
\begin{enumerate}
\item We summarize the latest progress of applying deep learning techniques to stock market prediction, especially those which only appear in the past three years.
\item We give a general workflow for stock market prediction, based on which the previous studies can be easily classified and summarized. And the future studies can refer to the previous work in each step of the workflow.
\item We pay a special attention to implementation and reproducibility, which is often neglected in similar surveys.
\item We point out several future directions, some of which are on-going and help the readers to catch up with the research frontiers.
\end{enumerate}

The rest of this survey is organize as follows: Section 2 presents related work; Section 3 gives an overview of the papers we cover; Section 4 describes the major findings in each step of the prediction workflow; Section 5 gives the discussion about implementation and reproducibility; Section 6 points up some possible future research directions; We conclude this survey in Section 7.

\section{Related Work}
    \label{sec:related}

Stock market prediction has been a research topic for a long time, and there are some review papers accompanied with the development and flourishment of deep learning methods prior to our work. While their focus could also be applications of deep learning methods, stock market prediction could only be one example of many financial problems in these previous surveys. In this section, we list some of them in a chronological order and discuss our motivation and unique perspectives.

Back to 2009, \cite{Atsalakis2009Surveying} surveys more than 100 related published articles that focus on neural and neuro-fuzzy techniques derived and applied to forecast stock markets, with the discussion of classifications of input data, forecasting methodology, performance evaluation and performance measures used. \cite{li2010applications} gives a survey on the application of artificial neural networks in forecasting financial market prices, including the forecast of stock prices, option pricing, exchange rates, banking and financial crisis. \cite{nikfarjam2010text} surveys some primary studies which implement text mining techniques to extract qualitative information about companies and use this information to predict the future behavior of stock prices based on how good or bad are the news about these companies.

\cite{aguilar2015genetic} presents a review of the application of evolutionary computation methods to solving financial problems, including the techniques of genetic algorithms, genetic programming, multi-objective evolutionary algorithms, learning classifier systems, co-evolutionary approaches, and estimation of distribution algorithms. \cite{cavalcante2016computational} gives an overview of the most important primary studies published from 2009 to 2015, which cover techniques for preprocessing and clustering of financial data, for forecasting future market movements, for mining financial text information, among others. \cite{tkavc2016artificial} provides a systematic overview of neural network applications in business between 1994 and 2015 and reveals that most of the research has aimed at financial distress and bankruptcy problems, stock price forecasting, and decision support, with special attention to classification tasks. Besides conventional multilayer feedforward network with gradient descent backpropagation, various hybrid networks have been developed in order to improve the performance of standard models.

More recently, ~\cite{xing2018natural} reviews the application of cutting-edge NLP techniques for financial forecasting, which would be concerned when text including the financial news or twitters is used as input for stock market prediction. ~\cite{rundo2019machine} covers a wider topic both in the machine learning techniques, which include deep learning, but also the field of quantitative finance from HFT trading systems to financial portfolio allocation and optimization systems. ~\cite{nti2019systematic} focuses on the fundamental and technical analysis, and find that support vector machine and artificial neural network are the most used machine learning techniques for stock market prediction. Based on its review of stock analysis, ~\cite{shah2019stock} points out some challenges and research opportunities, including issues of live testing, algorithmic trading, self-defeating, long-term predictions, and sentiment analysis on company filings.

Different from other related works that cover more papers from the computer science community, ~\cite{reschenhofer2019evaluation} reviews articles covered by the Social Sciences Citation Index in the category “Business, Finance” and gives more insight on economic significance. It also points out some problems in the existing literature, including unsuitable benchmarks, short evaluation periods, and nonoperational trading strategies.

Some latest reviews are trying to cover a wider range, e.g., ~\cite{shah2019stock} covers machine learning techniques applied to the prediction of financial market prices, and ~\cite{sezer2019financialreview} covers more financial instruments. However, our motivation is to catch up with the research trend of applying deep learning techniques, which have been proved to outperform traditional machine learning techniques, e.g., support vector machine in most of the publications, with only a few exceptions, e.g., ~\cite{ballings2015evaluating} finds that Random Forest is the top algorithm followed by Support Vector Machines, Kernel Factory, AdaBoost, Neural Networks, K-Nearest Neighbors and Logistic Regression, and ~\cite{ersan2019comparison} finds that K-Nearest Neighbor and Artificial Neural Network both outperform Support Vector Machines, but there is no obvious pros and cons between the performances of them. With the accumulation of historical prices and diverse input data types, e.g., financial news and twitter, we think the advantages of deep learning techniques would continue and it is necessary to keep updated with this trend for the future research.

Compared with ~\cite{sezer2019financialreview}, whose focus is deep learning for financial time series forecasting and a much longer time period (from 2005 to 2019 exactly), we focus on the recent progress in the past three years (2017-2019) and a narrower scope of stock price and market index prediction. For readers who are also interested in other financial instruments, e.g., commodity price, bond price, cryptocurrency price, etc., we would refer them to this work. We also care more about the implementation workflow and result reproducibility of previous studies, e.g., dataset and code availability, which is a problem that has drawn the attention from the AI researchers~\citep{gundersen2018state}. We would also pay more attention to the uniqueness of stock market prediction (or financial time series forecasting) from general time series prediction problems, e.g., the evaluation of profitability besides prediction accuracy.

\section{Overview}
    \label{sec:overview}
In this section, we give an overview of the papers we are going to review in this study. All the works are searched and collected from Google Scholar, with searching keywords such as deep learning, stock prediction, stock forecasting, etc. Most of the covered papers (115 out of 124) are published in the past three years (2017-2019). In total, we cover 56 journal papers, 58 conference papers and 10 preprint papers. These preprint papers are all from arXiv.org, which is a famous website for e-print archive and we cover these papers to keep updated with the latest progress. The top source journals \& conferences sorted by the number of papers we cover in this study are shown in Table~\ref{tab:source_journal} and Table~\ref{tab:source_conference}, respectively.

\begin{table}[!htb]
    \centering
    \caption{List of top source journals and the number of papers we cover in this study.}
    \label{tab:source_journal}
    \begin{tabular}{ll}
        \hline
        Journal Name & Paper Count \\
        \hline
        Expert Systems with Applications & 12 \\
        IEEE Access & 5 \\
        Neurocomputing & 3 \\
        Complexity & 2 \\
        Journal of Forecasting & 2 \\
        Knowledge-Based Systems & 2 \\
        Applied Soft Computing & 2 \\
        Mathematical Problems in Engineering & 2 \\
        PLOS ONE & 2 \\
        Others in total & 24 \\
        \hline
    \end{tabular}
\end{table}

\begin{table}[!htb]
    \centering
    \caption{List of top conferences and the number of papers we cover in this study.}
    \label{tab:source_conference}
    \begin{tabular}{p{0.6\textwidth}l}
        \hline
        Conference Name & Paper Count \\
        \hline
        International Joint Conferences on Artificial Intelligence (IJCAI) & 4 \\
        International Joint Conference on Neural Networks (IJCNN) & 4 \\
        Conference on Information and Knowledge Management (CIKM) & 3 \\
        International Conference on Neural Information Processing (ICONIP) & 3 \\
        Annual Meeting of the Association for Computational Linguistics (ACL) & 2 \\
        IEEE Symposium Series on Computational Intelligence (SSCI) & 2 \\
        Hawaii International Conference on System Sciences (HICSS) & 2 \\
        ACM SIGKDD Conference on Knowledge Discovery and Data Mining (KDD) & 2 \\
        IEEE International Conference on Tools with Artificial Intelligence (ICTAI) & 2 \\
        Others in total & 34 \\
        \hline
    \end{tabular}
\end{table}

In this study, the major focus would be the prediction of the close prices of individual stocks and market indexes. Some financial instrument whose price is bounded to the market index is also covered, e.g., some exchange-traded fund (ETF) or equity index futures that track the underlying market index. For intraday prediction, we would also cover mid-price prediction for limit order books. Other financial instruments are not mentioned in this study, e.g., bond price and cryptocurrency price. More specifically, if the target to predict the specific value of the prices, we classify it as a \textit{regression} problem, and if the target is to predict the price movement direction, e.g., going up or down, we classify it as a \textit{classification} problem. Most studies are considering the daily prediction (105 of 124) and only a few of them are considering the intraday prediction (18 of 124), e.g., 5-minute or hourly prediction. Only one of the 124 papers is considering both the daily and intraday situations~\citep{liu2018numerical}.

\begin{figure}[!htb]
    \centering
    \includegraphics[width=\textwidth]{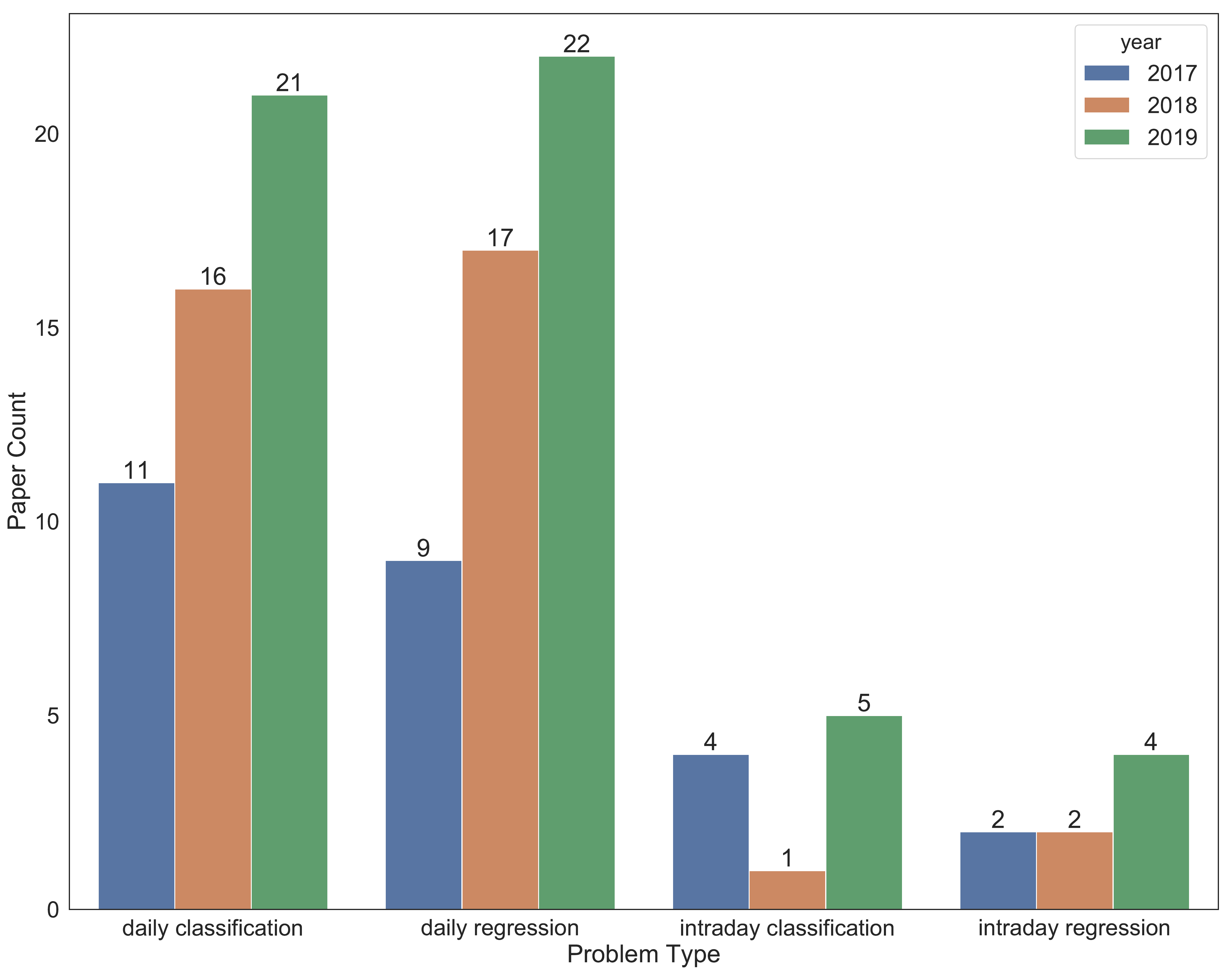}
    \caption{The paper count of different problem types.}
    \label{fig:problem}
\end{figure}

Based on the target output and frequency, the prediction problems can be classified into four types: daily classification (52 of 124), daily regression (54 of 124), intraday classification (8 of 124) and intraday regression (11 of 124). A detailed paper count of different prediction problem types is shown in Figure~\ref{fig:problem}. The reason behind this could be partially justified by the difficulty of collecting the corresponding data. The daily historical prices and news titles are easier to collect and process for research, while the intraday data is very limited in the academia. We would further discuss the data availability in Section~\ref{sec:implementation}.

\begin{figure}[!htb]
    \centering
    \includegraphics[width=\textwidth]{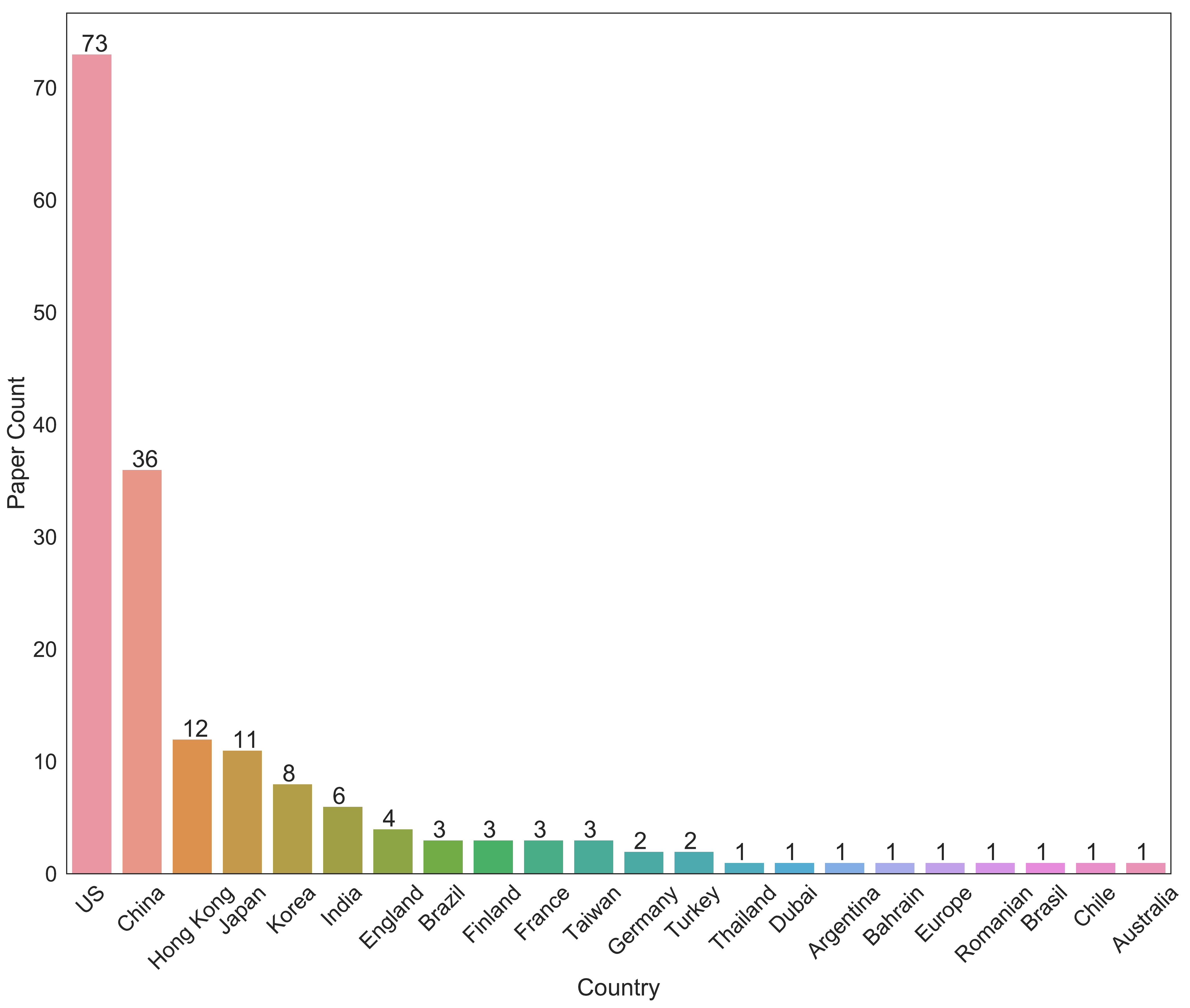}
    \caption{The paper count from different markets.}
    \label{fig:fig1}
\end{figure}

Surveyed markets as well as the most famous stock market index in these markets are shown in the Table~\ref{tab:market_index}. The paper count of different surveyed markets is shown in Figure~\ref{fig:fig1}~\footnote{An exception of using Europe for~\cite{ballings2015evaluating}, in which 5767 publicly listed European companies are covered.}. Most of the studies would focus on one market, while some of them would evaluate their models on multiple markets~\footnote{Due to the differences of trading rules, we list mainland China, Hong Kong and Taiwan separately.}. Both mature markets (e.g., US) and emerging markets (e.g., China) are gaining a lot of attention from the research community in the past three years. 

\begin{center}
\begin{longtable}{p{0.1\textwidth}p{0.2\textwidth}p{0.6\textwidth}}
\caption{List of surveyed markets and stock indexes.} \label{tab:market_index} \\
\hline Country & Index & Description \\ \hline 
\endfirsthead

\multicolumn{3}{c}%
{{\bfseries \tablename\ \thetable{} -- continued from previous page}} \\
\hline Country & Index & Description \\ \hline 
\endhead

\hline \multicolumn{3}{r}{{Continued on next page}} \\ \hline
\endfoot

\hline
\endlastfoot

US & S\&P 500 & Index of 505 common stocks issued by 500 large-cap companies \\
US & Dow Jones Industrial Average & Index of 30 major companies \\
US & NASDAQ Composite & Index of common companies in NASDAQ stock market \\
US & NYSE Composite & Index of common companies in New York Stock Exchange \\
US & RUSSEL 2000 & Index of bottom 2,000 stocks in the Russell 3000 Index \\
China & SSE Composite & Index of common companies in Shanghai Stock Exchange \\
China & CSI 300 & Index of top 300 stocks in Shanghai and Shenzhen stock exchanges \\
Hong Kong & HSI & Hang Seng Index of the largest companies in Hong Kong Exchange \\
Japan & Nikkei 225 & Index of 225 large companies in Tokyo Stock Exchange \\
Korea & Korea Composite & Index of common companies in Korea Stock Exchange \\
India & BSE 30 & Index of 30 companies exist in Bombay Stock Exchange \\
India & NIFTI 50 & Index of 50 companies exist in National Stock Exchange \\
England & FTSE 100 & Index of 100 companies in London Stock Exchange \\
Brazil & IBOV & Bovespa Index of 60 stocks \\
France & CAC 40 & Index of 40 stocks most significant stocks in Euronext Paris \\
Germany & DAX & Index of 30 major German companies in Frankfurt Stock Exchange \\
Turkey & BIST 100 & Index of 100 stocks in Borsa Istanbul Stock Exchange \\
Argentina & MER & Merval Index in Buenos Aires Stock Exchange \\
Bahrain & BAX & Bahrain All Share Index of 42 stocks \\
Chile & IPSA & Ipsa Index of 40 most liquid stocks \\
Australia & All Ordinaries & Index of 500 largest companies in Australian Securities Exchange \\
\end{longtable}
\end{center}

\section{Prediction Workflow}
    \label{sec:workflow}

Given different combinations of data sources, previous studies explored the use of deep learning models to predict stock market price/movement. In this section, we summarize the previous studies in a general workflow with four steps that most of the studies follow: Raw Data, Data Processing, Prediction Model and Model Evaluation. In this section, we would discuss each step separately and reveal a general approach that the future work can easily reproduce.

\subsection{Raw Data}
The first step of predicting is to collect proper data as the basis. It could be the intrinsic historical prices with the assumption that history repeats itself, or the extrinsic data sources that affect the stock market. In the efficient-market hypothesis, asset prices already reflect all available information. However, in practice many researchers do not agree with this conclusion, thus many different extrinsic sources of data are used for stock market prediction, e.g., \cite{weng2017stock} compares the usage of market data, technical indicators, Wikipedia traffic, Google news counts, and generated features, and ~\cite{liu2019combining} covers market data, fundamental data, knowledge graph, and news.

\subsubsection{Data Types}
In this part, we categories the raw data that are commonly used for stock market prediction into seven types:

\begin{itemize}
    \item Market data: market data includes all trading activities that happen in a stock market, e.g., open/high/low/close prices, trading volume, etc. It is used as both input features (e.g., the historical prices in a look-back window) and prediction target (e.g., the close price of the next day).
    \item Text data: text data refer to the text contributed by individuals, e.g., social media, news, web searches, etc. As a type of alternative data, these data are hard to collect and process, but may provide useful information that is not included in market data. Sentiment analysis can be applied on these text data and produce a sentiment factor (e.g., positive, neural, or negative) that can be further used for prediction.
    \item Macroeconomics data: macroeconomics data reflects the economic circumstances of a particular country, region or sector, e.g., Consumer Price Index (CPI), Gross domestic product (GDP), etc. These indicators are related with the stock market in the sense that they indicate how healthy the overall stock market is and can provide confirmation as to the quality of a stock market advance or decline.
    \item Knowledge graph data: there are some kind of relationship between different companies and different markets, e.g., the movement of stocks in the same sector may be affected by the same news. Powered by the recently developed graph neural networks, the knowledge graph data from open sources such as FreeBase~\citep{bollacker2008freebase} and Wikidata~\footnote{\url{https://www.wikidata.org/}} can now be used to improve the prediction performance.
    \item Image data: inspired by the success of convolutional neural networks in 2D image processing, e.g., classification and object detection, candlestick charts are used as input images for stock prediction. While satellite and CCTV images or videos are used to monitor the situation of companies and may be helpful for stock price prediction, they are never used in the surveyed papers because of the prohibitive cost of collection and the potential privacy leakage risk.
    \item Fundamental data: the most common type of fundamental data is the accounting data, which is reported quarterly, e.g., assets, liabilities, etc. It is less used in studies with deep learning models because the low frequency of reporting and also the inaccuracies of the reporting date, e.g., the fundamental data published is indexed by the last date included in the report and precedes the date of the release, which brings in a risk of using future information.
    \item Analytics data: analytics data refers to the data that can be extracted from reports (e.g., recommendation for selling or buying a stock) that are provided by investment banks and research firms, who make an in-depth analysis of companies' business models, activities, competitions, etc. These reports provide valuation information, while they may be costly and shared among different consumers, who all want to use this information to make a profit.
\end{itemize}

\begin{figure}[!htb]
    \centering
    \includegraphics[width=\textwidth]{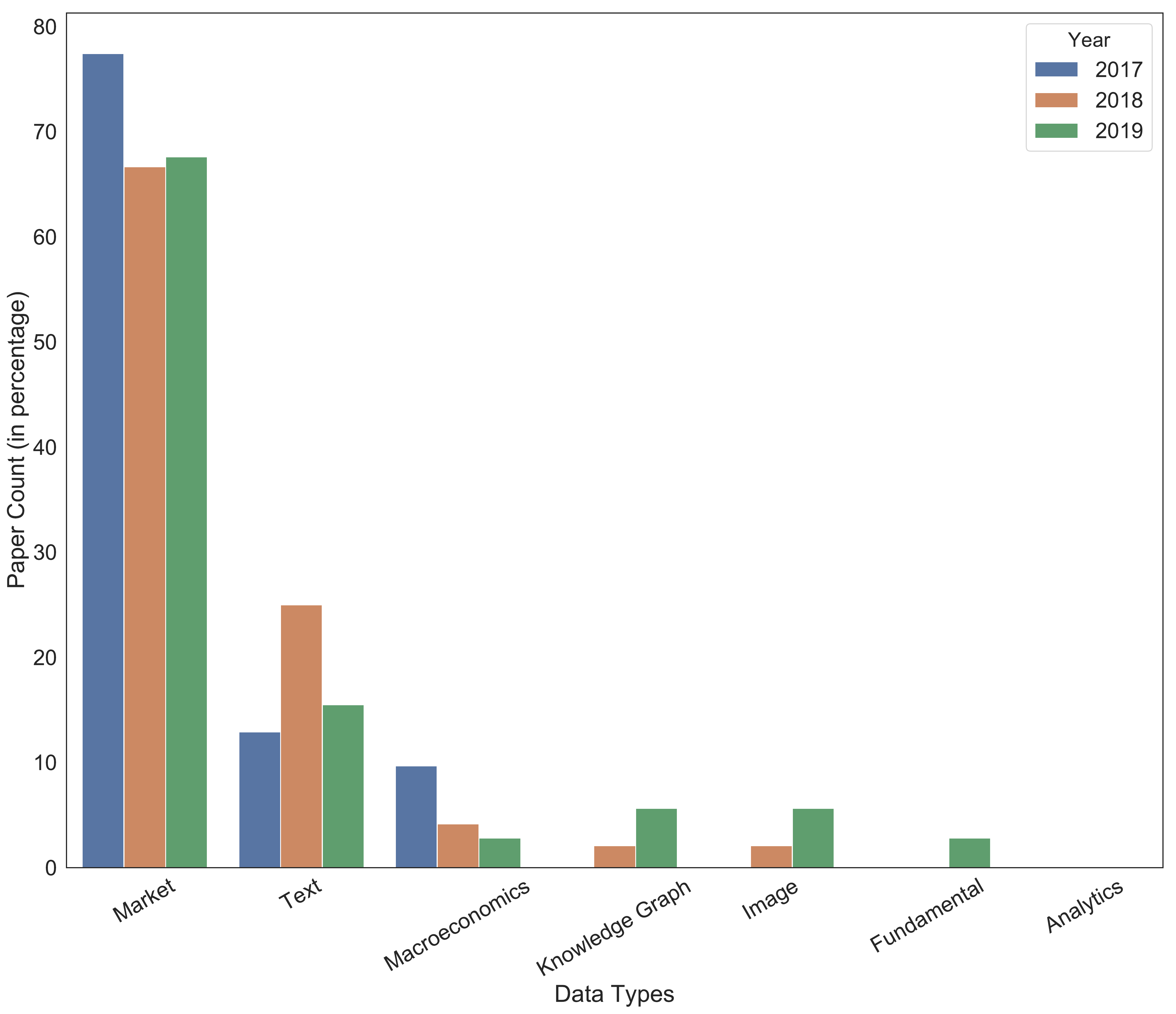}
    \caption{The usage of different raw data types.}
    \label{fig:types}
\end{figure}

Different types or raw data are accompanied with different levels of difficulty for obtaining and processing, and the usage of different data types is shown in Figure~\ref{fig:types}. For deep learning models, a huge amount of input data is necessary for the training of a complex neural network model. In this case, market data is the best choice and used for the most as it provides the largest amount of data sample, while the other data types usually have a smaller size. Text data is used for the second most, with the popularity of social media and online news website and the easier use of web crawlers to get the text data. An extreme case is the analytics data, which is never used in the surveyed studies, because of both the data sparsity and the high cost to access.

There is also a trend in Figure~\ref{fig:types} that more diverse data types are used in 2018 and 2019, compared to the studies in 2017. It indicates the fact that it is harder to get a better prediction result based on only the market data. It also reflects the development of new tools so that new types of data can be used for prediction, e.g., graph neural networks for knowledge graph data.

\subsubsection{Data Length}
To evaluate the performance of different models, historical data is necessary for evaluation. However, there is a tradeoff of choosing the data length. A short time period of data is not sufficient to show the effective and has a higher risk of overfitting, while a long time period takes the risk of traversing different market styles and present out-of-dated results. Besides, the data availability and cost are factors that needs to be taken into consideration when choosing the data length.

\begin{figure}[!htb]
    \centering
    \includegraphics[width=\textwidth]{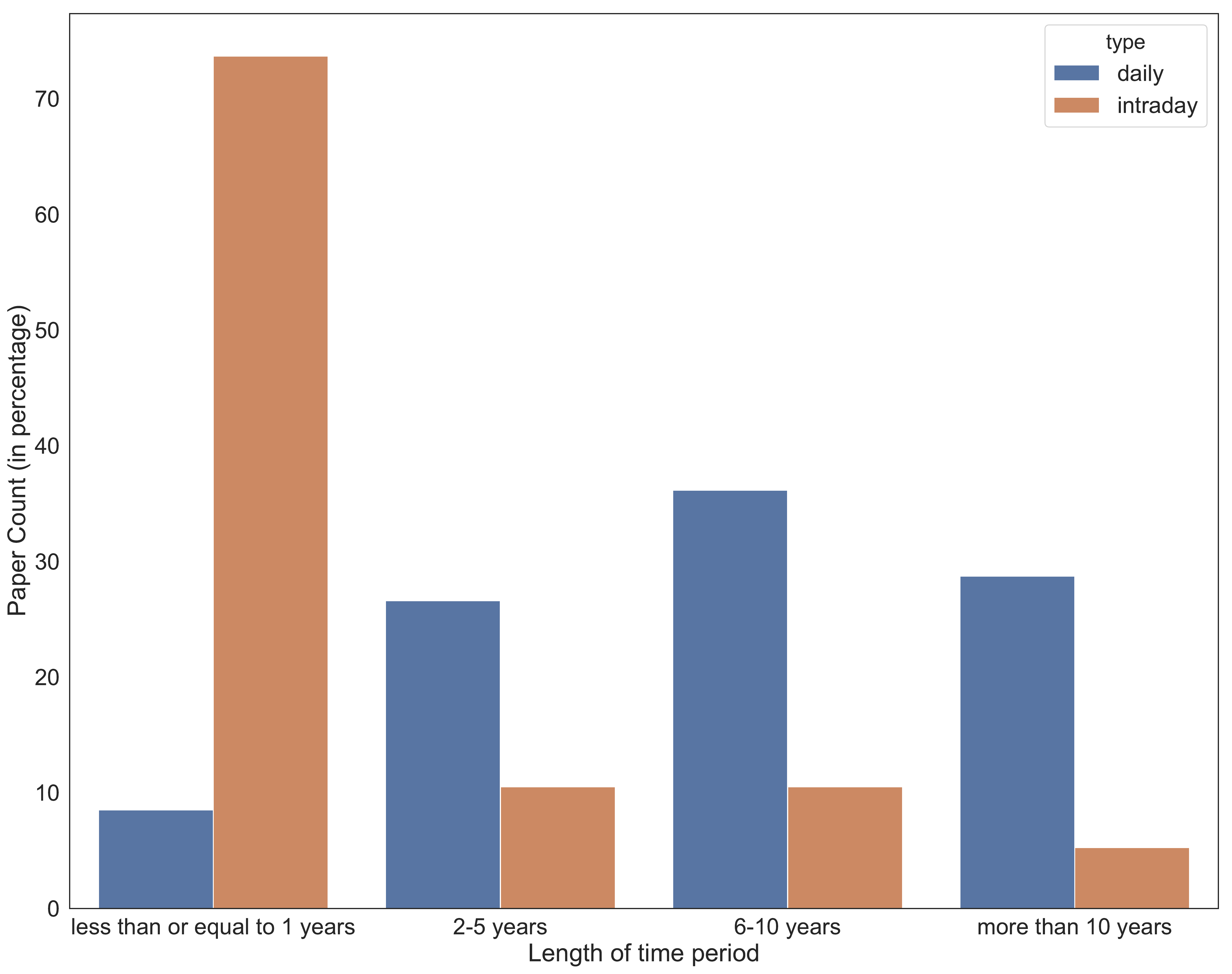}
    \caption{The distribution of data length.}
    \label{fig:time}
\end{figure}

The distribution of time periods of data used in the surveyed papers is shown in Figure~\ref{fig:time}. It is more expensive to get intraday data with a good quality and most of the previous studies involving intraday prediction would use a time period less than one year.

For a single prediction, \textit{lag} is used to denote the time length of the input data to be used by the model, e.g., in the daily prediction, a lag of 30 days means the data in the past 30 days are used to build the input features. For technical indicators, lag is often set as an input parameters and vary a lot in previous studies from 2 to 252 time periods. Correspondingly, \textit{horizon} is used to denote the time length of the future to be predicted by the model. Most of the studies focus on a short-term prediction horizon, e.g., one day or five minutes, with only a few exceptions for a longer horizon such as five days or ten days.

\subsection{Data Processing}
\subsubsection{Missing Data Imputation}
The problem of missing data is not as severe as in other domains, e.g., sensor data, because the market data is more reliable and well supported and maintained by the trading markets. However, to align multiple types of data with different sampling frequencies, e.g., market data and fundamental data, the data with a lower sampling frequency should be inserted in a forward way by propagating the last valid observation forward to next valid, to avoid data leakage of the future information.

\subsubsection{Denoising}
With many irrational behaviors in the stock trading process, the market data is filled with noise, which may misrepresent the trend of price change and misguide the prediction. As a signal processing technique, wavelet transform has been used to eliminate noise in stock price time series~\citep{bao2017deep, liang2019lstm}. Another approach to eliminate noisy data in~\cite{sun2017stacked} is the use of the kNN-classifiers, based on two training sets with different labels in a data preparation layer.

\subsubsection{Feature Extraction}
For machine learning models, feature engineering is the process of extracting input features from raw data based on domain knowledge. Combined with raw data, these handcrafted features are used as input for the prediction models and can substantially boost machine learning model performance.

For market data, technical analysis is a feature extraction approach that builds various indicators for forecasting the direction of prices based on historical prices and volumes, \textit{e.g.}, moving average, or moving average convergence/divergence (MACD). These technical indicators can be further used to design simple trading strategies. Technical indicators are also used to build image inputs, \textit{e.g.}, 15 different technical indicators with a 15-days periods are used to construct a 15 $\times$ 15 sized 2D images in \cite{sezer2018algorithmic}.

While the feature extraction techniques represented by technical analysis for market data have been used and validated for many years, the tools for extracting features from text data have made a greater progress in the past few years, owing to the various deep learning models developed for natural language processing. Before the popularity of machine learning models, the bag-of-words (BoW) model~\citep{harris1954distributional} is used as a representation of text that describes the occurrence of words within a document. In recent years, machine learning and deep learning models show an improved performance for word embedding. Given the sequence of words, the word2vec model~\citep{mikolov2013efficient}, which are shallow and two-layer neural networks, can be used to embed each of these words in a vector of real numbers and has been used in~\cite{liu2019transformer, liu2019anticipating, lee2017predict}. Global Vectors for Word Representation (GloVe)~\citep{pennington2014glove} is another word embedding method proposed by Stanford University, in which each of the word vectors has a dimension of 50, and has been used in~\cite{tang2018stock}.

Stock markets are highly affected by some public events, which can be extracted from online news data and used as input features. ~\cite{ding2015deep} uses a neural tensor network to learn event embeddings for representing news documents. \cite{hu2018listening} uses a news embedding layer to encode each news into a news vector. ~\cite{wang2019ean} uses a convolution neural network (CNN) layer to extract salient features from transformed event representations. 

From sentiment aspect, the text data can be further analyzed and a sentiment vector can depict each word, which may present the positive or negative opinions for the future direction of stock prices. For sentiment analysis, \cite{jin2019stock} uses CNN and ~\cite{mohan2019stock} uses Natural Language Toolkit (NLTK)~\citep{loper2002nltk}. ~\cite{minh2018deep} even proposes a sentiment Stock2Vec embedding model trained on both the stock news and the Harvard IV-4 psychological dictionary, which may not be directly related to stock market prediction.

Off-the-shelf commercial software is also available for linguistic features and sentiment analysis. For example, ~\cite{kumar2019predicting} employs Linguistic Inquiry and Word Count (LIWC)~\citep{tausczik2010psychological} to find out the linguistic features in the news articles, which includes the text analysis module along with a group of built-in dictionaries to count the percentage of words reflecting different emotions, thinking styles, social concerns, and even parts of speech.

\begin{figure}[!htb]
    \centering
    \includegraphics[width=\textwidth]{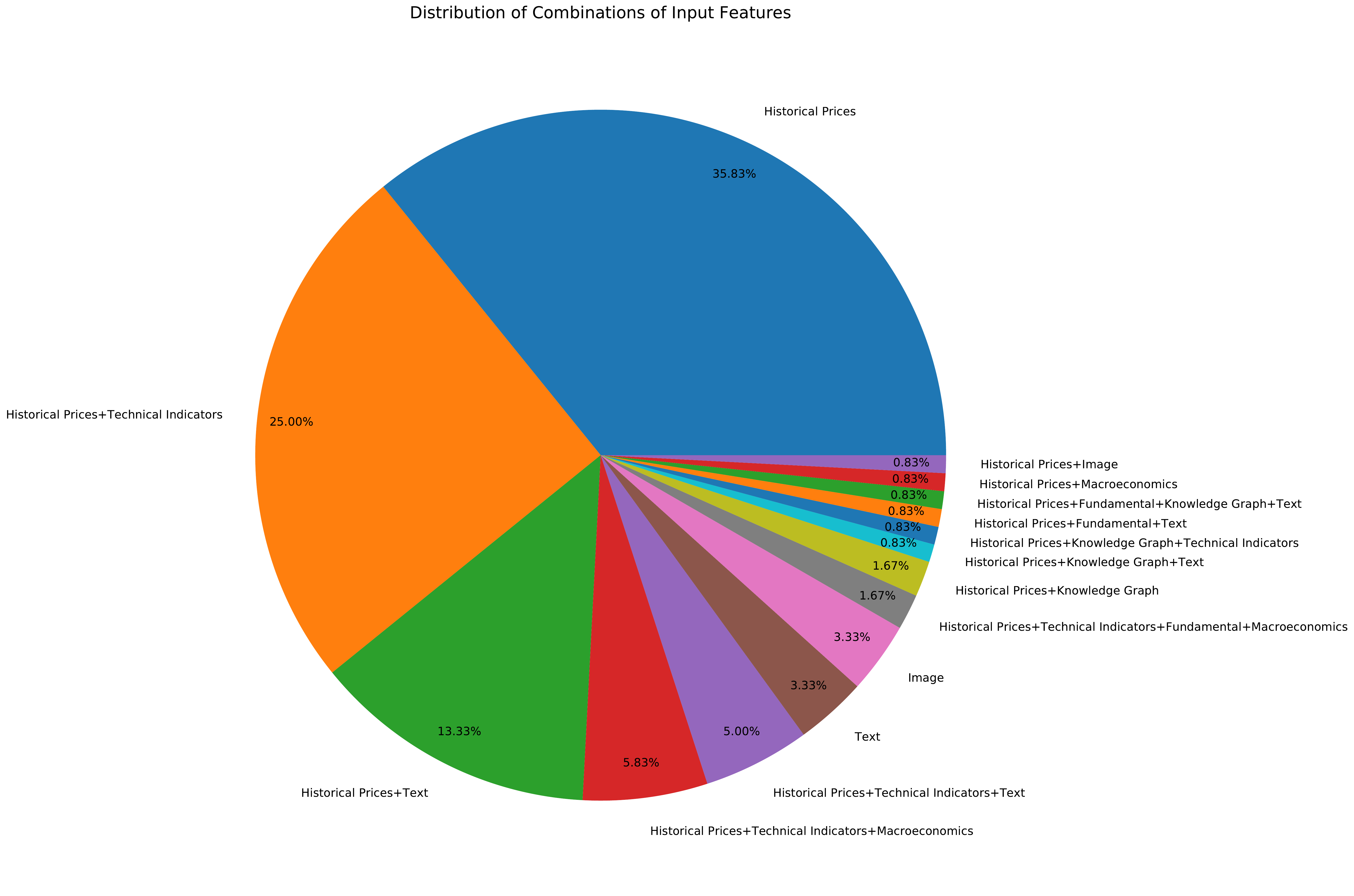}
    \caption{Distribution of combinations of input features.}
    \label{fig:input_features}
\end{figure}

For knowledge graph data used more recently, the TransE model~\citep{bordes2013translating} is a computationally efficient predictive model that satisfactorily represents a one-to-one type of relationship and has been used in~\cite{liu2019anticipating}.

Based on the input raw data and extracted features, we show the distribution of different combinations of input features in Figure~\ref{fig:input_features} and the detailed article lists in Table~\ref{tab:input_features}. From Figure~\ref{fig:input_features}, historical prices and technical indicators are the most commonly used input features and followed by text and macroeconomics data. This could be explained by the easier accessing and processing of market data than other data types.

\begin{center}
\begin{longtable}{p{0.4\textwidth}p{0.6\textwidth}}
\caption{Article Lists of combinations of input features.}
\label{tab:input_features} \\
\hline Combination & Article List \\ \hline 
\endfirsthead

\multicolumn{2}{c}%
{{\bfseries \tablename\ \thetable{} -- continued from previous page}} \\
\hline Combination & Article List \\ \hline 
\endhead

\hline \multicolumn{2}{r}{{Continued on next page}} \\ \hline
\endfoot

\hline
\endlastfoot

Historical Prices & ~\cite{chen2019hybrid, ding2019study, zhou2019emd2fnn, nguyen2019predicting, yang2017stock, li2017combining, li2017time, sachdeva2019effective, nikoustock, tsantekidis2017using, liu2019non, qin2017dual, siami2019comparative, guang2019multi, zhao2017time, althelaya2018stock, baek2018modaugnet, liang2019lstm, fischer2018deep, pang2018innovative, tran2018temporal, tsantekidis2017forecasting, chen2018artificial, wang2019clvsa, zhang2017stock, karathanasopoulos2019forecasting, hollis2018comparison, kim2019financial, cao2019stock, selvin2017stock, araujo2019deep, wang2015forecasting, zhang2019stock, zhang2017data, wu2018adaboost, long2019deep, cao2019financial, hossain2018hybrid, eapen2019novel, zhan2018stock, lei2018wavelet, chong2017deep, siami2018comparison} \\
Historical Prices + Technical Indicators &  ~\cite{assis2018restricted, cheng2018applied, nelson2017stock, gunduz2017intraday, sanboon2019deep, chen2019exploring, stoean2019deep, al2019forecasting, yu2019ceam, li2019multitask, gao2018improving, chung2018genetic, yan2018financial, sethia2019application, sun2019exploiting, zhang2019deeplob, chen2017double, zhou2018stock, borovkova2019ensemble, chen2019a, feng2019enhancing, sun2017stacked, yang2018multi, ticknor2013bayesian, song2019study, goccken2016integrating, singh2017stock, patel2015predicting, liu2017stock, merello2019ensemble} \\
Historical Prices + Text & ~\cite{jin2019stock, li2019dp, liu2019transformer, liu2018numerical, wang2019ean, xu2018stock, matsubara2018stock, tang2018stock, huang2018tensor, wu2018hybrid, kumar2019predicting, li2017sentiment, tang2019learning, huynh2017new, mohan2019stock, hu2018predicting} \\
Historical Prices + Technical Indicators + Macroeconomics & ~\cite{dingli2017financial, de2013applying, zhong2017forecasting, bao2017deep, xie2018recurrent, hoseinzade2019cnnpred, hoseinzade2019u} \\
Historical Prices + Technical Indicators + Text & ~\cite{vargas2017deep, liu2019anticipating, oncharoen2018deep, minh2018deep, lee2017predict, chen2018leveraging} \\
Text & ~\cite{liu2018hierarchical, hu2018listening, ding2015deep, ding2014using} \\
Image & ~\cite{sezer2018algorithmic, sim2019deep, lee2019global, sezer2019financial} \\
Historical Prices + Technical Indicators + Fundamental + Macroeconomics & ~\cite{ballings2015evaluating, niaki2013forecasting} \\
Historical Prices + Knowledge Graph & ~\cite{kim2019hats, chen2018incorporating} \\
Historical Prices + Knowledge Graph + Text & ~\cite{deng2019knowledge} \\
Historical Prices + Knowledge Graph + Technical Indicators & ~\cite{matsunaga2019exploring} \\
Historical Prices + Fundamental data + Text & ~\cite{tan2019tensor} \\
Historical Prices + Fundamental + Knowledge Graph + Text & ~\cite{liu2019combining} \\
Historical Prices + Macroeconomics & ~\cite{jiang2018cross} \\
Historical Prices + Image & ~\cite{kim2019forecasting} \\
\end{longtable}
\end{center}


\subsubsection{Dimensionality Reduction}
It is possible that many features are highly correlated with each other, e.g., the technical indicators which are all calculated from historical open/high/low/close prices and volume. To alleviate the corresponding problem of deep learning model's overfitting, dimensionality reduction for the input features has been adopted as a preprocessing technique for stock market prediction.

Principal component analysis (PCA) is a commonly used transformation technique that uses Singular Value Decomposition of the input data to project it to a lower dimensional space and has been used in~\cite{gao2018improving, zhang2019deeplob, zhong2017forecasting, wang2015forecasting, singh2017stock, chong2017deep}. \cite{zhong2017forecasting} even gives a comparison between different versions of PCA, and finds that the PCA-ANN model give a slightly higher prediction accuracy for the daily direction of SPY for next day, compared to the use of fuzzy robust principal component analysis (FRPCA) and kernel-based principal component analysis (KPCA).

For dimensionality reduction, the other options include independent components analysis (ICA)~\citep{sethia2019application}, autoencoder~\citep{chong2017deep}, restricted Boltzmann machine~\citep{chong2017deep}, empirical mode decomposition (EMD)~\citep{cao2019financial, zhou2019emd2fnn}, and sub-mode coordinate algorithm (SMC)~\citep{huang2018tensor}. \cite{huang2018tensor} first utilizes tensor to integrate the multi-sourced data and further proposes an improved SMC model to reduce the variance of their subspace in each dimension produced by the tensor decomposition.



Feature selection is another way of dimensionality reduction, by choosing only a subset of input features. Chi-square method~\citep{zheng2004feature} and maximum relevance and minimum redundancy (MRMR)~\citep{peng2005feature} are two common used feature selection techniques. Chi-square method decides whether a categorical predictor variable and the target class variable are independent or not. High chi-squared values indicate the dependence of the target variable on the predictor variable. Minimum redundancy maximum relevance uses a heuristic to minimize redundancy while maximizing relevance to select promising features for both continuous and discrete inputs, through F-statistic values. Chi-square method is used in~\cite{gunduz2017intraday, kumar2019predicting} and maximum relevance and minimum redundancy is used in~\cite{kumar2019predicting}.

Other options for feature selection include rough set attribute reduction (RSAR)~\citep{lei2018wavelet}, autocorrelation function (ACF) and partial correlation function (PCF)~\citep{wu2018adaboost}, the analysis of variance (ANOVA)~\citep{niaki2013forecasting}, and maximal information coefficient feature selection (MICFS)~\citep{yang2018multi}.










\subsubsection{Feature Normalization \& Standardization}
Given different input features with varying scales, feature normalization and standardization are used to guarantee that some machine learning models can work and also help to improve the model's training speed and performance. Feature normalization refers to the process of rescaling the input feature by the minimum and range, to make all the values lie between 0 and 1~\citep{wang2015forecasting, gunduz2017intraday, li2017time, singh2017stock, althelaya2018evaluation, althelaya2018stock, baek2018modaugnet, chen2018artificial, chung2018genetic, gao2018improving, hossain2018hybrid, hu2018predicting, minh2018deep, pang2018innovative, tang2018stock, xie2018recurrent, yang2018multi, zhan2018stock, al2019forecasting, araujo2019deep, cao2019financial, cao2019stock, ding2019study, lee2019global, li2019dp, sachdeva2019effective, sethia2019application, tang2019learning}, or between -1 and 1~\citep{ticknor2013bayesian, zhang2017stock, sezer2018algorithmic}. Feature standardization means subtracting a measure of location and dividing by a measure of scale, e.g., the z-score method that subtracts the mean and divides by the standard deviation~\citep{tsantekidis2017forecasting, tsantekidis2017using, zhang2019deeplob, li2019multitask}.





\subsubsection{Data Split}
For evaluation of different prediction models, in-sample/out-of-sample split or train/validation/test split of data samples is commonly used in machine learning and deep learning fields. The model is trained with the training or in-sample data set, the hyper-parameters is fine-tuned on the validation data set optional, and the final performance is evaluated on the test or out-of-sample data set. k-fold cross validation is further used to split the dataset into k consecutive folds, and k-1 folds is used as the training set, while the last fold is then used as a test set.

\begin{figure*}[!htb]
\centering
\subfigure[][]{
    \includegraphics[width=0.4\textwidth]{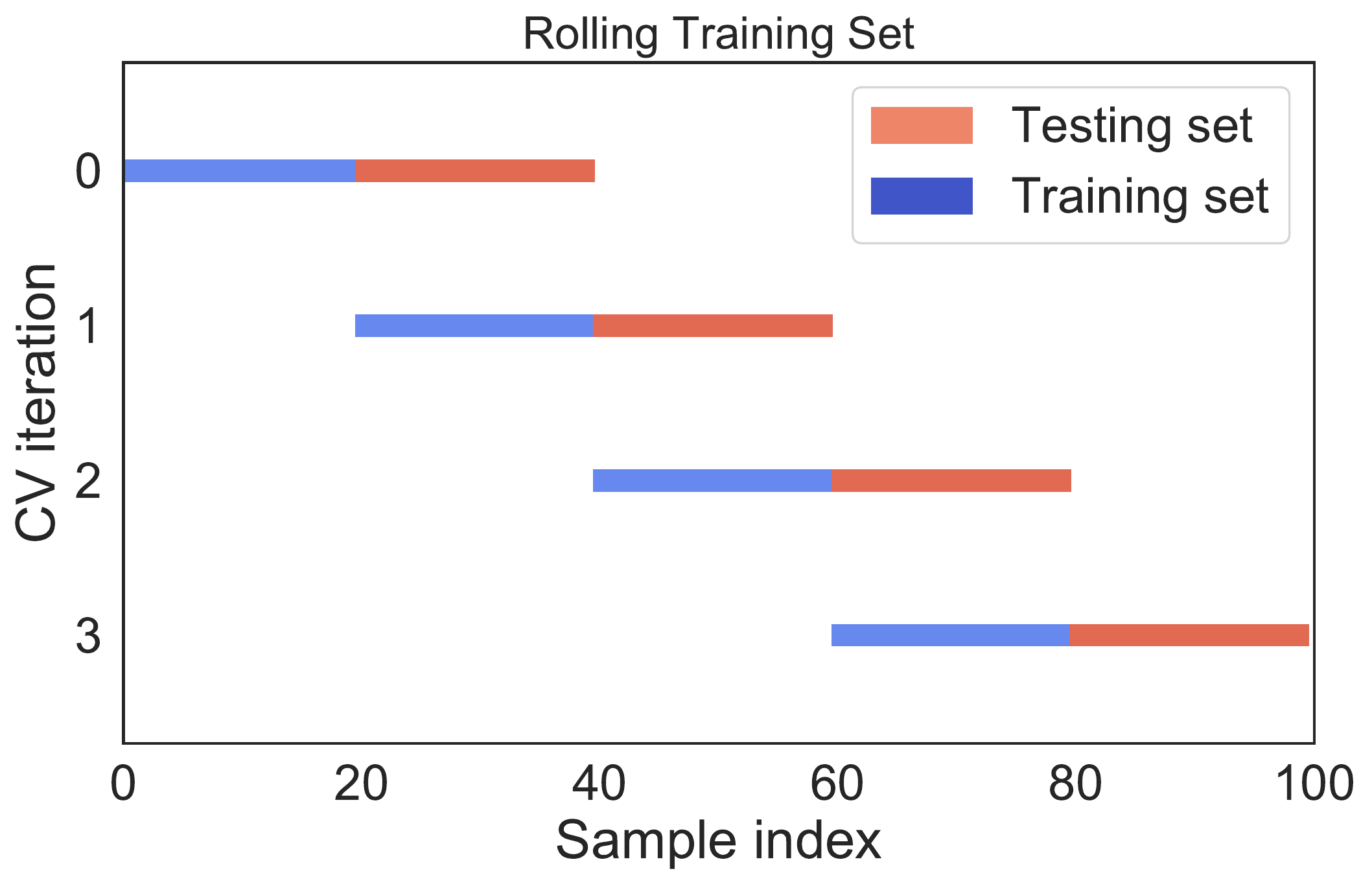}
    \label{fig:rolling_training}
}
\subfigure[][]{
    \includegraphics[width=0.4\textwidth]{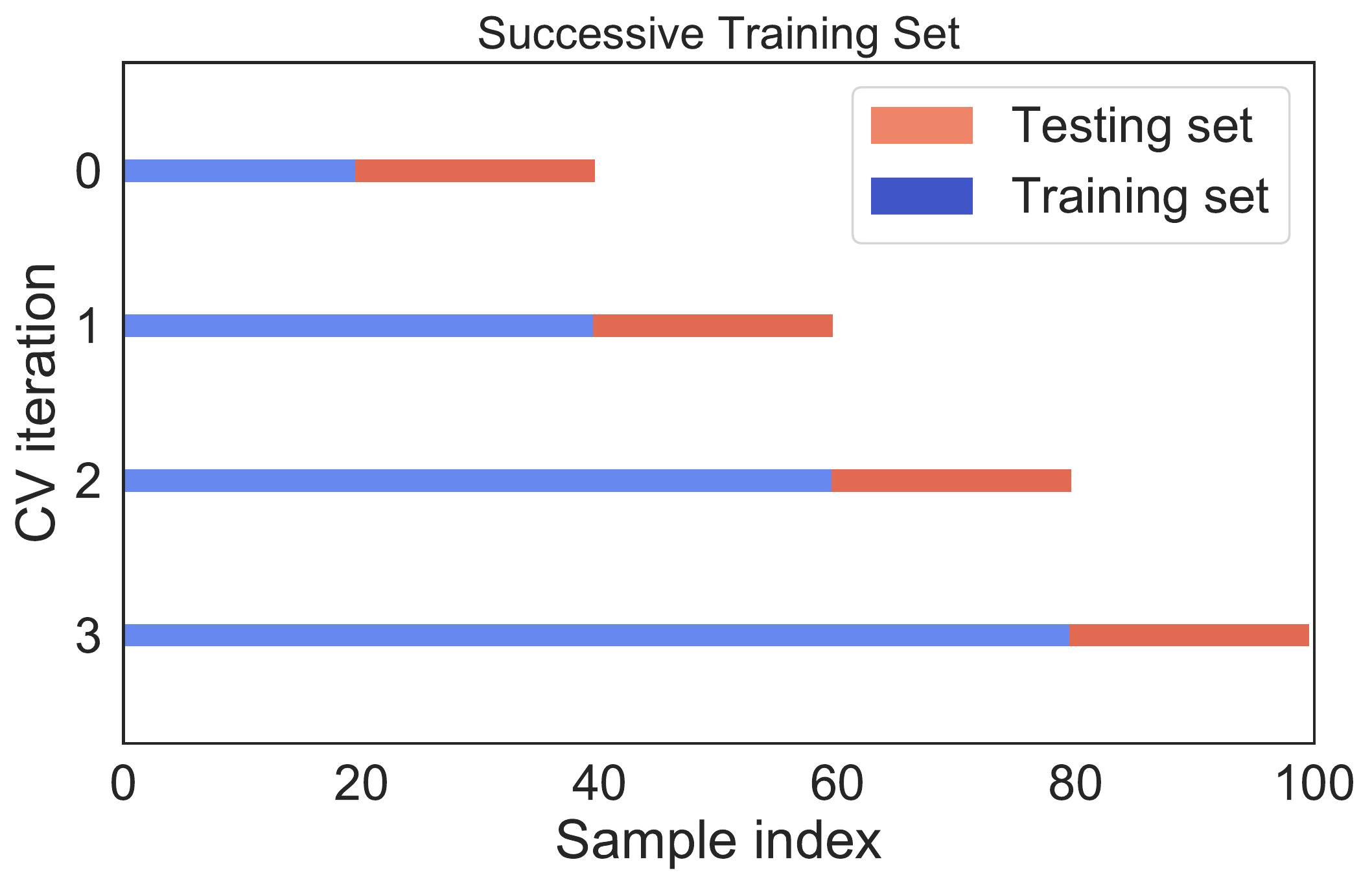}
    \label{fig:successive_training}
}
\caption[]{The example or rolling training-validation-test data splits.
    \subref{fig:rolling_training} Rolling training set;
    \subref{fig:successive_training} Successive training set.
}
\label{fig:rolling}
\end{figure*}

As a special case, train-validation-test split with a rolling (or sliding, moving, walk-forward) window is also often used for time series tasks including stock prediction~\citep{bao2017deep, nelson2017stock, fischer2018deep, gao2018improving, zhou2018stock, nguyen2019novel, kim2019hats, sun2019exploiting, wang2019clvsa}. The process of a rolling train-validation-test split is shown in Figure~\ref{fig:rolling_training}, where only the latest part of data samples are used for a new training round of prediction models. Another variant is to use successive training sets, which are union set of the rolling training set that come before them, as shown in Figure~\ref{fig:successive_training}.

\subsubsection{Data Augmentation}
Data augmentation techniques have been widely used for image classification and object detection tasks and proved to effectively enhance the classification and detection performance. However, it is less used for time series tasks including stock prediction, even though the size of stock price time series is not comparable to the size of public image datasets, which usually have millions of sample and even more in recent years.

There are still a few works that explore the usage of data augmentation. \cite{zhang2017data} firstly clusters different stocks based on their retracement probability density function and combines all the day-wise information of the same stock cluster as enlarged training data. In the ModAugNet framework proposed in \cite{baek2018modaugnet}, the authors choose 10 companies' stock that are highly correlated to the stock index and augment the data samples by using the combinations of 10 companies taken 5 at a time in an overfitting prevention LSTM module, before feeding the data samples to a prediction LSTM module for stock market index prediction.



\subsection{Prediction Model}
Most of the prediction models belong to a supervised learning approach, when training set is used for the training and test set is used for evaluation. Only a few of the studies use semi-supervised learning when the labels are not available in the feature extraction step. We further classify the various prediction models into three types: standard models and their variants, hybrid models, and other models. For standard models, three families of deep learning models, namely, feedforward neural network, convolutional neural network and recurrent neural network, are used a lot. And we category the use of generative adversarial network, transfer learning, and reinforcement learning into other models. These models only appear in recent years and are still in an early stage of being applied for stock market prediction.

In this part, we are focusing on the usage of different types of deep learning models, instead of diving into the details of each model. For a more detailed introduction to deep learning models, we refer the readers to~\cite{goodfellow2016deep}. The abbreviations of machine learning and deep learning methods are shown in Table~\ref{tab:abbr}.

\begin{center}
\begin{longtable}{p{0.2\textwidth}p{0.8\textwidth}}
\caption{Abbreviations of machine learning and deep learning methods.}
\label{tab:abbr} \\
\hline Abbreviation & Full Name \\ \hline 
\endfirsthead

\multicolumn{2}{c}%
{{\bfseries \tablename\ \thetable{} -- continued from previous page}} \\
\hline Abbreviation & Full Name\\ \hline 
\endhead

\hline \multicolumn{2}{r}{{Continued on next page}} \\ \hline
\endfoot

\hline
\endlastfoot

\hline
\multicolumn{2}{l}{Preprocessing Techniques} \\
\hline
BoF & Bag-of-feature \\
BoW & Bag-of-words \\
CEAM & Cycle Embeddings with Attention Mechanism \\
CEEMDAN & Complete Ensemble Empirical Mode Decomposition with Adaptive Noise \\
EMD & Empirical Mode Decomposition \\
MACD & Moving Average Convergence / Divergence \\
MOM & Momentum \\
N-BoF & Neural Bag-of-Feature \\
PCA & Principal Components Analysis \\
RS & Rough Set \\
RSI & Relative Strength Index \\
SMA & Simple Moving Average \\
SMC & Sub-mode Coordinate Algorithm \\
WT & Wavelet Transform \\
(2D)2PCA & 2-Directional 2- Dimensional Principal Component Analysis \\

\hline
\multicolumn{2}{l}{Optimization Algorithms} \\
\hline

HM & Harmony Memory \\
GA & Genetic Algorithm \\
ISCA & Improved Sine Cosine Algorithm \\
GWO & Grey Wolf Optimizer \\
PSO & particle swarm optimization \\
WOA & Whale Optimization Algorithm \\
SCA & Sine Cosine Algorithm \\

\hline
\multicolumn{2}{l}{Linear Models} \\
\hline
AR & Autoregressive \\
ARMA & Autoregressive Moving Average\\
ARIMA & Autoregressive Integrated Moving Average \\
EMA & Estimated Moving Average \\
GARCH & Generalized Autoregressive Conditional Heteroskedasticity \\
LDA & Linear Discriminant Analysis \\
LR & Linear Regression \\
PLR & Piecewise Linear Regression \\
MA & Moving Average \\
MR & Mean Reversion \\
MCSDA & Multilinear Class-specific Discriminant Analysis \\
MDA & Multilinear Discriminant Analysis \\
MTR & Multilinear Time-series Regression \\
WMTR & Weighted Multilinear Time-series Regression \\

\hline
\multicolumn{2}{l}{Machine Learning Models} \\
\hline
AB & AdaBoost \\
ANFIS & Adaptive Neuro-fuzzy Inference System \\
kNN & k-Nearest neighbor \\
Lasso & Lasso Regression \\
Logit & Logistics Regression \\
MKL & Multiple Kernel Learning \\
MRFs & Markov Random Fields \\
RF & Random Forest \\
Ridge & Ridge Regression \\
SVM & Support Vector Machine \\
SVR & Support Vector Regression \\

\hline
\multicolumn{2}{l}{Deep Learning Models} \\
\hline

AE & Autoencoder \\
ANN & Artificial Neural Network \\
BGRU & Bidirectional GRU \\
BiLSTM & bi-directional long short-term memory \\
BPNN & Backpropagation Neural Network \\
CapTE & Capsule network based Transformer Encoder \\
CF-DA-RNN & DA-RNN with cycle value \\
CH-RNN & Cross-modal Attention based Hybrid Recurrent Neural Network \\
CNN & Convolutional Neural Network \\
DA-RNN & Dual-Stage Attention-Based RNN \\
DBN & Deep Belief Network \\
DEM & Dilation-erosion Model \\
DGM & Deep Neural Generative Model \\
DIDLNN & Deep Increasing–decreasing-linear Neural Network \\
DMN & Dendrite Morphological Neuron \\
DNN & Deep Neural Network \\
ELM & Extreme Learning Machines \\
eLSTM & Tensor-based Event-LSTM \\
EMD2FNN & Empirical Mode Decomposition and Factorization Machine based Neural Network \\
EMD2NN & Empirical Mode Decomposition based Neural Network \\
FDNN & Fuzzy Deep Neural Network \\
FFNN & Feedforward Neural Network \\
FNN & Factorization Machine based Neural Network \\
GAN & Generative Adversarial Network \\
GCNN & Graph Convolutional Neural network \\
GNN & Graph Neural Network \\
GRU & Gated Recurrent Unit \\
HAN & Discriminative Deep Neural Network with Hierarchical Attention \\
HCAN & Hierarchical Complementary Attention Network \\
HGAN & Hierarchical Graph Attention Network \\
KDTCN & Knowledge-Driven Temporal Convolutional Network \\
LNNN & Linear and Nonlinear Neural Network \\
LSTM & Long Short Term Memory \\
MAFN & Multi-head Attention Fusion Network \\
MGU & Minimal Gated Unit \\
MLP & Multilayer Perceptron \\
MS-CNN & Multi-Scale CNN \\
MSTD-RCNN & Multi-Scale Temporal Dependent Recurrent Convolutional Neural Network \\
NARXT & Nonlinear Autoregressive Neural Network with Exogenous Takens Inputs \\
NN & Neural Network \\
PELMNN & Prediction Evolutionary Levenberg-marquardt Neural Networks \\
RBFNN & Radial Basis Function Neural Network \\
RBM & Restricted Boltzmann Machines \\
RNN & Recurrent Neural Network \\
SAEs & Stacked Autoencoders \\
SFM & State Frequency Memory \\
SRCGUs & Selective Recurrent Neural Networks with Random Connectivity Gated Unit \\
STNN & Stochastic Time Effective Function Neural Network \\
TABL & Temporal Attention Augmented Bilinear Layer \\
TGC & Temporal Graph Convolution \\
TSLDA & Generative Topic Model Jointly Learning Topics and Sentiments \\
WMN & Wavelet Neural Network \\
WDBPNN & Wavelet De-noising based BPNN \\
1D CNN & One-dimensional Convolutional Neural Networks \\

\end{longtable}
\end{center}

Some of the earlier work use ANNs as their prediction models and study the effect of different combinations of input features~\citep{niaki2013forecasting, de2013applying, zhong2017forecasting}. In this survey, we use ANN to refer to the neural networks which only have one or zero hidden layers, and DNN to refer to those which have two or more hidden layers. The list of standard models or their variants is shown in Table~\ref{tab:standard}. We organize the standard models into three major types:

\begin{itemize}
    \item \textit{Feedforward neural network (FFNN)}. It is the simplest type of artificial neural network wherein connections between the nodes do not form a cycle. An artificial neural networks (ANN) are learning models inspired by biological neural networks, and the neuron in an ANN consists of an aggregation function which calculates the sum of the inputs, and an activation function which generates the outputs. An autoencoder (AE) is a subset of ANN which has the same number of nodes in the input and output layers. When ANN has two or more hidden layers, we denote it as deep neural network (DNN) is this survey. We also category the following models into this family because they share a similar structure: backpropagation neural network (BPNN), multilayer perceptron (MLP), extreme learning machines (ELM) where the parameters of hidden nodes need not be tuned, deep increasing–decreasing-linear neural network (IDLNN) where each layer is composed of a set of increasing–decreasing-linear processing units~\citep{araujo2019deep}, stochastic time effective function neural network (STEFNN)~\citep{wang2015forecasting}, radial basis function network (RBFN) that uses radial basis functions as activation functions.
    \item \textit{Convolutional neural network (CNN)}. Designed for processing two-dimensional images, each group of neurons, which is also called a filter, performs a convolution operation to a different region of the input image and the neurons share the same weights, which reduces the number of parameters compared to the densely connected feedforward neural network. Pooling operations, \textit{e.g.}, max pooling, are used to reduce the original size and can be used for multiple times, until the final output is concatenated to a dense layer. Powered by the parallel processing ability of graphics processing unit (GPU), the training of CNN has been shortened and CNN has achieved an astonishing performance for image related tasks and competitions. By reducing the convolutional and pooling operations to a single temporal dimension, 1D CNN is proposed for time series classification and prediction, e.g., \cite{deng2019knowledge} uses a 1-D fully-convolutional network (FCN) architecture, where each hidden layer has the same length as the input layer, and zero padding is added to keep subsequent layers the same length as previous ones.
    \item \textit{Recurrent neural network (RNN)}. Compared with feedforward neural network, recurrent neural network is an artificial neural network wherein connections between the nodes form a cycle along a temporal sequence, which helps it to exhibit temporal dynamic behavior. However, normal RNNs are bothered by the vanishing gradient problem in practice, when the gradients of some of the weights start to shrink or enlarge if the network is unfolded too many times. Long short-term memory (LSTM) networks are RNNs that solve the vanishing gradient problem, where the hidden layer is replaced by recurrent gates called forget gates. Gated recurrent unit (GRU) is another RNN that uses forget gates, but has fewer parameters than LSTM. Bi-directional RNN are RNNs that connect two hidden layers of opposite directions to the same output. Both bi-directional LSTM (BiLSTM) and bi-directional GRU (BGRU) have been used for stock market prediction.
\end{itemize}

While standard models perform well at early stages of research, their variants are further developed to improve the prediction performance. One approach is to use stacked models, where neural network sub-models are embedded in a larger stacking ensemble model for training and prediction. Another approach is to introduce the attention mechanism~\citep{treisman1980feature} into recurrent neural network models, in which attention is a generalized pooling method with bias alignment over inputs.

There are also some types that we list separately:
\begin{itemize}
    \item Restricted Boltzmann machine (RBM) is a generative stochastic artificial neural network that can learn a probability distribution over its set of inputs. And a deep belief network (DBN) can be defined as a stack of RBMs. DBNs have been used in~\cite{li2017time, karathanasopoulos2019forecasting} for stock prediction.
    \item Sequence to sequence (seq2seq) model is based on the encoder-decoder architecture to generate a sequence output for a sequence input, in which both the encoder and the decoder use recurrent neural networks. Seq2seq model has been used in~\cite{liu2018numerical} for stock prediction.
\end{itemize}

While our focus in this survey is not the linear models or the traditional machine learning models, they are often used as baselines for comparison with deep learning models.

Some often used linear prediction models are as follows:
\begin{itemize}
    \item \textit{Linear regression (LR)}. Linear regression is a classical linear model that tries to fit the relationship between the predicted target and the input variables with a linear model, in which the parameters can be learned in the least squares approach.
    \item \textit{Autoregressive integrated moving average (ARIMA).} ARIMA is a generalization of the autoregressive moving average (ARMA) model, which describes a weakly stationary stochastic process with two parts, namely, the autoregression (AR) and the moving average (MA). Compared with ARMA, ARIMA is capable of dealing with non-stationary time series, by introducing an initial differencing step, which is referred as the integrated part in the model.
    \item \textit{Generalized autoregressive conditional heteroskedasticity (GARCH).} GARCH is also a generation of the autoregressive conditional heteroscedasticity (ARCH) model, which describes the error variance as a function of the actual sizes of the previous time periods' error terms. Instead of using AR model in ARCH, GARCH assumes an ARMA model for the error variance, which generalizes ARCH.
\end{itemize}

Similarly, some often used machine learning models are as follows:
\begin{itemize}
    \item \textit{Logistics regression (Logit)}. Logistics regression can be seen as a generalized linear model, in which a logistic function is used to model the probabilities of a binary target of being 0 or 1. It is suitable for the classification of price movements, e.g., going up or down.
    \item \textit{Support vector machine/regression (SVM/SVR)}. Support vector machine is a classical and powerful tool for classification with a good theoretical performance guarantee and has been widely adopted before the popularity of deep learning models. SVM tries to learn a hyperplane to distinguish the training samples that maximize distance of the decision boundary from training samples. Combine with the kernel trick, which maps the input training samples into high-dimensional feature spaces, SVM can efficiently perform non-linear classification tasks. Support vector regression is the regression version of SVM.
    \item \textit{k-nearest neighbor (kNN)}. kNN is a non-parametric model for both classification and regression, in which the output is the class most common or the average of the values among $k$ nearest neighbors. A useful technique is to assign weights to the neighbors when combing their contributions.
\end{itemize}

Given the predicted movement direction or prices, a long-short strategy can be further designed to perform trading based on the prediction model, e.g., if the predicted direction is going up, long it, otherwise short it. A simple baseline is the Buy\&Hold Strategy, which buys the asset at the beginning and hold it to the end of the testing period, without any further buying or selling operations~\citep{niaki2013forecasting, sezer2018algorithmic}.

Technical indicators are also often used for designing baseline trading strategies, e.g., momentum strategy, which is introduced in~\cite{jegadeesh1993returns}, is a simple strategy of buying winners and selling losers. MACD~\citep{appel2007understanding} consists of the MACD line and the signal line, and the most common MACD strategy buys the stock when the MACD line crosses above the signal line and sells the stock when the MACD line crosses below the signal line~\citep{lee2019global}. In our surveyed papers, RSI (14 days, 70-30) and SMA (50 days) are used to design baseline trading strategies~\citep{sezer2018algorithmic}.

\begin{center}
\begin{longtable}{p{0.3\textwidth}p{0.3\textwidth}p{0.4\textwidth}}
\caption{List of standard models or their variants.} \label{tab:standard} \\
\hline Article & Prediction Model & Baselines \\ \hline 
\endfirsthead

\multicolumn{3}{c}%
{{\bfseries \tablename\ \thetable{} -- continued from previous page}} \\
\hline Article & Prediction Model & Baselines \\ \hline 
\endhead

\hline \multicolumn{3}{r}{{Continued on next page}} \\ \hline
\endfoot

\hline
\endlastfoot

\hline
\multicolumn{3}{l}{Feedforward Neural Network Type} \\
\hline
\cite{de2013applying} & ANN & N/A \\
\cite{niaki2013forecasting} & ANN & Logit, Buy\&Hold Strategy \\
\cite{ticknor2013bayesian} & DNN & ARIMA \\
\cite{ding2014using} & DNN & SVM \\
\cite{ding2015deep} & DNN & BoW+SVM, structed event tuple+ANN \\
\cite{wang2015forecasting} & PCA+STNN & BPNN, PCA+BPNN, STNN \\
\cite{goccken2016integrating} & HM-ANN & ANN, GA-ANN \\
\cite{chen2017double} & DNN & ARMA-GARCH, ARMAX-GARCH, ANN \\
\cite{chong2017deep} & DNN & AR \\
\cite{singh2017stock} & (2D)2PCA+RBFNN & RBFNN, RNN \\
\cite{sun2017stacked} & Stacked Denoising Autoencoder & SVM, Logit, ANN \\
\cite{zhong2017forecasting} & PCA+ANN & N/A \\
\cite{hu2018predicting} & ISCA–BPNN & BPNN, GWO-BPNN, PSO-BPNN, WOA-BPNN, SCA-BPNN \\
\cite{araujo2019deep} & DIDLNN & ARIMA, SVM, MLP, LNNN~\citep{yolcu2013new}, DEM~\citep{araujo2011class}, DMN~\citep{zamora2017dendrite}, NARXT~\citep{menezes2008long}, PELMNN~\citep{asadi2012hybridization} \\
\cite{song2019study} & DNN & DNN \\

\hline
\multicolumn{3}{l}{Convolutional Neural Network Type} \\
\hline
\cite{dingli2017financial} & CNN & Logit, SVM \\
\cite{gunduz2017intraday} & CNN & Logit \\
\cite{selvin2017stock} & 1-D CNN & RNN, LSTM, ARIMA \\
\cite{tsantekidis2017forecasting} & CNN & SVM, MLP \\
\cite{sezer2018algorithmic} & CNN & Buy\&Hold Strategy, RSI (14 days, 70-30), SMA (50 days), LSTM and MLP \\
\cite{yang2018multi} & multichannel CNN & SVM, ANN, CNN \\
\cite{hoseinzade2019cnnpred} & CNN & PCA+ANN~\citep{zhong2017forecasting}, ANN~\citep{kara2011predicting}, CNN~\citep{gunduz2017intraday} \\
\cite{cao2019stock} & 1-D CNN & DNN, SVM \\
\cite{deng2019knowledge} & KDTCN & ARIMA, LSTM, CNN, TCN \\
\cite{sezer2019financial} & CNN & Buy\&Hold \\
\cite{sim2019deep} & CNN & ANN, SVM \\

\hline
\multicolumn{3}{l}{Recurrent Neural Network Type} \\
\hline
\cite{huynh2017new} & BGRU & LSTM, GRU, DNN~\citep{ding2014using}, DNN~\citep{peng2016leverage} \\
\cite{li2017combining} & WT+LSTM & LSTM \\
\cite{li2017sentiment} & LSTM & SVM \\
\cite{nelson2017stock} & LSTM & MLP, RF \\
\cite{tsantekidis2017using} & LSTM & SVM, MLP \\
\cite{zhang2017data} & GRU & SVM \\
\cite{zhang2017stock} & SFM & AR, LSTM \\
\cite{zhao2017time} & LSTM & RNN, SVM, RF, AB \\
\cite{althelaya2018evaluation} & bidirectional and stacked LSTM & ANN, LSTM \\
\cite{althelaya2018stock} & Stacked LSTM & BiLSTM, BGRU, Stacked GRU, MLP \\
\cite{baek2018modaugnet} & LSTM & DNN, RNN \\
\cite{cheng2018applied} & Attention LSTM & N/A \\
\cite{chung2018genetic} & GA+LSTM & N/A \\
\cite{fischer2018deep} & LSTM & RF, DNN, Logit \\
\cite{gao2018improving} & LSTM & MA, EMA, ARMA, GARCH, SVM, FFNN, and LSTM \\
\cite{hollis2018comparison} & Attention LSTM & LSTM \\
\cite{huang2018tensor} & SMC+LSTM & SVM, PCA+SVM, TeSIA~\citep{li2016tensor}, SMC+TeSIA \\
\cite{jiang2018cross} & RNN with attention & RNN \\
\cite{liu2018hierarchical} & HCAN & BoW~\citep{joachims1998text}, FastText~\citep{joulin2017bag}, Structured-Event~\citep{ding2014using}, IAN~\citep{ma2017interactive} \\
\cite{minh2018deep} & two-stream GRU & LSTM, GRU \\
\cite{siami2018comparison} & LSTM & ARIMA \\
\cite{xie2018recurrent} & LSTM & WT+SAEs+LSTM~\citep{bao2017deep} \\
\cite{xu2018stock} & StockNet & Random Guess, ARIMA, RF, TSLDA~\citep{nguyen2015topic}, HAN~\citep{hu2018listening} \\
\cite{yan2018financial} & WT+LSTM & MLP, SVM, kNN \\
\cite{borovkova2019ensemble} & stacked LSTM & Lasso, Ridge \\
\cite{cao2019financial} & EMD+LSTM, CEEMDAN+LSTM & LSTM, SVM, MLP \\
\cite{chen2019exploring} & Attention LSTM & LSTM \\
\cite{chen2019hybrid} & EMD-Attention LSTM & MLP, LSTM, EMD-LSTM, Attention LSTM \\
\cite{ding2019study} & LSTM & RNN, LSTM \\
\cite{feng2019enhancing} & Adversarial Attentive LSTM & MOM, MR, StockNet~\citep{xu2018stock}, LSTM~\citep{nelson2017stock}, Attentive RNN ~\citep{qin2017dual} \\
\cite{liang2019lstm} & WT+LSTM & LSTM \\
\cite{liu2019combining} & GRU & N/A \\
\cite{kim2019financial} & weighted LSTM with Attention & MLP, 1D CNN, stacked LSTM, LSTM with attention \\
\cite{mohan2019stock} & LSTM & ARIMA, Facebook Prophet, LSTM \\
\cite{nguyen2019predicting} & Dynamic LSTM & LSTM \\
\cite{nikoustock} & LSTM & ANN, SVM, RF \\
\cite{sachdeva2019effective} & LSTM & N/A \\
\cite{sanboon2019deep} & LSTM & SVM, MLP, DT, RF, Logit, kNN \\
\cite{sethia2019application} & LSTM & GRU, ANN, SVM \\
\cite{siami2019comparative} & BiLSTM & LSTM, ARIMA \\
\cite{tan2019tensor} & eLSTM & SVM, DT, ANN, LSTM, AZFin Text~\citep{Schumaker2009Textual}, TeSIA~\citep{li2016tensor} \\
~\cite{tran2018temporal} & TABL & Ridge, FFNN, LDA, MDA, MTR, WMTR~\citep{tran2017tensor}, MCSDA~\citep{tran2017multilinear}, BoF, N-BoF~\citep{passalis2017time}, SVM, MLP, CNN~\citep{tsantekidis2017forecasting}, LSTM~\citep{tsantekidis2017using} \\
\cite{wang2019ean} & BiLSTM & Random guess, ARIMA, SVM, MLP, HAN~\citep{hu2018listening} \\

\hline
\multicolumn{3}{l}{Other Types} \\
\hline
\cite{li2017time} & DBN & N/A \\
\cite{matsubara2018stock} & DGM & SVM, MLP \\
\cite{liu2018numerical} & seq2seq model with attention & AZFin Text~\citep{Schumaker2009Textual}, DL4S~\citep{akita2016deep}, DA-RNN~\citep{qin2017dual}, MKL, ELM~\citep{li2016empirical} \\
\cite{karathanasopoulos2019forecasting} & DBN & MACD, ARMA \\
\end{longtable}
\end{center}

We further categories the hybrids into two classes, namely, the hybrid models between deep learning models and traditional models, and the hybrid models between different deep learning models.

The list of hybrid models between deep learning models and traditional models is shown in Table~\ref{tab:hybrid_dl_ml}. \cite{li2019dp} formulates a sentiment-ARMA model to incorporate the news articles as hidden information and designs a LSTM-based DNN, which consists of three components, namely, LSTM, VADER model and differential privacy mechanism that integrates different news sources. To deal with strong noise, \cite{liu2017stock} uses weak ANNs to get some information without over-fitting and get better results by combining the weak results together using optimized bagging. \cite{wu2018adaboost} uses AdaBoost algorithm to generate both training samples and ensemble weights for each LSTM predictor and the final prediction results are the combination of all the LSTM predictors with ensemble weights.




\begin{center}
\begin{longtable}{p{0.3\textwidth}p{0.3\textwidth}p{0.4\textwidth}}
\caption{List of hybrid models between deep learning models and traditional models.} \label{tab:hybrid_dl_ml} \\
\hline Article & Prediction Model & Baselines \\ \hline 
\endfirsthead

\multicolumn{3}{c}%
{{\bfseries \tablename\ \thetable{} -- continued from previous page}} \\
\hline Article & Prediction Model & Baselines \\ \hline 
\endhead

\hline \multicolumn{3}{r}{{Continued on next page}} \\ \hline
\endfoot

\hline
\endlastfoot
\cite{patel2015predicting} & SVR–ANN & SVR-RF, SVR-SVR \\
\cite{liu2017stock} & Bagging+ANN & SVM, ANN, GA-ANN, RF \\
\cite{yang2017stock} & Bagging+ANN & N/A \\
\cite{assis2018restricted} & RBM+SVM & SVM \\
\cite{chen2018leveraging} & RNN+AdaBoost & MLP, SVR, RNN \\
\cite{lei2018wavelet} & 2RS-WNN & BP-NN, RBF-NN, ANFIS-NN, SVM, WNN, RS-WNN \\
\cite{wu2018adaboost} & AB-LSTM & ARIMA, MLP, SVR, ELM, LSTM, AB-MLP, AB-SVR, AB-ELM \\
\cite{chen2019a} &  PLR+CNN+Dual Attention Mechanism based Encoder-Decoder & SVR, LSTM, CNN, LSTM\_CNN~\citep{lin2017hybrid}, TPM\_NC \\
\cite{li2019dp} & LSTM+ARIMA & LSTM \\
\cite{li2019multitask} & RNN with high-order MRFs & LSTM, attention based LSTM Encoder-Decoder~\citep{bahdanau2014neural}, DA-RNN~\citep{qin2017dual} \\
\cite{sun2019exploiting} & ARMA-GARCH-NN & DNN, LSTM \\
\cite{zhou2019emd2fnn} & EMD2FNN & ANN, FNN, EMD2NN, WDBPNN~\citep{wang2011forecasting}, Long-short Strategy \\
\end{longtable}
\end{center}

The list of hybrid models between different deep learning models is shown in Table~\ref{tab:hybrid_dls}. Two popular combinations are the combination of CNN and RNN structures and the combination of different RNNs.

For the former case, TreNet~\citep{lin2017hybrid} hybrids LSTM and CNN for stock trend classification. \cite{zhang2019deeplob} proposes a deep learning model, comprising three main building blocks that include a standard convolutional layer, an Inception Module and a LSTM layer. \cite{guang2019multi} uses convolutional units to extract multi-scale features that precisely describe the state of the financial market and capture temporal, and uses a recurrent neural network to capture the temporal dependency and complementary across different scales.

For the latter case, \cite{al2019forecasting} proposes a forecasting model, using a combination of LSTM autoencoder and stacked LSTM network. \cite{wang2019clvsa} proposes a hybrid model, consisting of stochastic recurrent networks, the sequence-to-sequence architecture, the self- and inter-attention mechanism, and convolutional LSTM units. \cite{liu2019non} proposes an elective Recurrent Neural Networks with Random Connectivity Gated Unit (SRCGUs) that train random connectivity LSTMs, GRUs and MGUs simultaneously.

\begin{center}
\begin{longtable}{p{0.3\textwidth}p{0.3\textwidth}p{0.4\textwidth}}
\caption{List of hybrid models between different deep learning models.} \label{tab:hybrid_dls} \\
\hline Article & Prediction Model & Baselines \\ \hline 
\endfirsthead

\multicolumn{3}{c}%
{{\bfseries \tablename\ \thetable{} -- continued from previous page}} \\
\hline Article & Prediction Model & Baselines \\ \hline 
\endhead

\hline \multicolumn{3}{r}{{Continued on next page}} \\ \hline
\endfoot

\hline
\endlastfoot
\cite{bao2017deep} & WT+SAEs+LSTM & WT+LSTM, LSTM, RNN \\
\cite{lee2017predict} & RNN+CNN & LSTM \\
\cite{qin2017dual} & DA-RNN & ARIMA, NARX RNN~\citep{diaconescu2008use}, Encoder-Decoder~\citep{cho2014learning}, attention based LSTM Encoder-Decoder~\citep{bahdanau2014neural} \\
\cite{vargas2017deep} & CNN+LSTM & DNN~\citep{ding2014using, ding2015deep}, CNN~\citep{ding2015deep} \\
\cite{chen2018artificial} & DNN+AE+RBM & ANN, ELM, RBFNN \\
\cite{hossain2018hybrid} & LSTM+GRU & MLP, RNN, CNN \\
\cite{hu2018listening} & Hybrid Attention Networks & RF, MLP, GRU, BGRU, Temporal-Attention-RNN, News-Attention-RNN \\
\cite{oncharoen2018deep} & CNN+LSTM & N/A \\
\cite{pang2018innovative} & AE+LSTM & DBN, MLP, DBN+MLP \\
\cite{tang2018stock} & CNN+LSTM & LSTM, FFNN, CNN \\
\cite{wu2018hybrid} & CH-RNN & DA-RNN~\citep{qin2017dual} \\
\cite{zhan2018stock} & 1D CNN+LSTM & LSTM, GRU \\
\cite{al2019forecasting} & LSTM AE+stacked LSTM & LSTM, MLP \\
\cite{eapen2019novel} & CNN+BiLSTM & SVR \\
\cite{guang2019multi} & MSTD-RCNN & SVM, RF, FDNN, TreNet~\citep{lin2017hybrid}, SFM RNN~\citep{hu2017state}, MS-CNN~\citep{cui2016multi}\\
\cite{jin2019stock} & CNN+LSTM+Attention & LSTM \\
\cite{kim2019forecasting} & CNN+LSTM & CNN, LSTM \\
\cite{liu2019non} & SRCGUs & MGUs, GRUs and LSTMs \\
\cite{long2019deep} & CNN+RNN & RNN, LSTM, CNN, SVM, Logit, RF, LR \\
\cite{tang2019learning} & MAFN & MA, RF, XGBoost, SVR, Adversarial Attentive LSTM~\citep{feng2019enhancing}, HAN~\citep{hu2018listening}, StockNet~\citep{xu2018stock} \\
\cite{wang2019clvsa} & A Convolutional LSTM Based Variational Seq2seq Model with Attention & CNN, LSTM, seq2seq model with attention \\
\cite{yu2019ceam} & CEAM+DA-RNN & DA-RNN~\citep{qin2017dual}, CF-DA-RNN \\
\cite{zhang2019deeplob} & CNN+LSTM & SVM, MLP, CNN~\citep{tsantekidis2017forecasting} \\
\end{longtable}
\end{center}

We list other types of models in Table~\ref{tab:other}. We category five types of models in this part, which have not been fully explored for stock market prediction but already show some promising results.

\begin{itemize}
    \item \textit{Generative adversarial network (GAN).} GAN is introduced by~\cite{goodfellow2014generative}, in which a discriminative net $D$ learns to distinguish whether a given data instance is real or not, and a generative net $G$ learns to confuse $D$ by generating high quality fake data. This game between $G$ and $D$ would lead to a Nash equilibrium. Since the introduction of GAN, it has been applied in multiple image-related tasks, especially for image generation and enhancement, and generates a large family of variants. Inspired the success of GANs, ~\cite{zhou2018stock} proposes a generic GAN framework employing LSTM and CNN for adversarial training to predict high-frequency stock market.
    \item \textit{Graph neural network (GNN).} GNN is designed to utilize graph-structured data, thus capable of utilizing the network structure to incorporate the interconnectivity of the market and make better predictions, compared to relying solely on the historical stock prices of each individual company or on hand-crafted features~\citep{matsunaga2019exploring}. \cite{chen2018incorporating} first constructs a graph including 3,024 listed companies based on investment facts from real market, then learns a distributed representation for each company via node embedding methods applied on the graph, and applies three-layer graph convolutional networks to predict. \cite{kim2019hats} uses LSTM for the individual stock prediction task and GRU for the index movement prediction task where an additional graph pooling layer is needed.
    \item \textit{Capsule Network.} Different from the method of CNNs and RNNs, the capsule network increases the weights of similar information through its dynamic routing, which is proposed by~\cite{sabour2017dynamic} and displaces the pooling operation used in conventional convolution neural network. \cite{liu2019transformer} is the first to introduce the capsule network for the problem of stock movements prediction based on social media and show that the capsule network is effective for this task.
    \item \textit{Reinforcement learning.} Unlike supervised learning, reinforcement learning trains an agent to choose the optimal action given a current state, with the goal to maximize cumulative rewards in the training process. Reinforcement learning can be applied for stock prediction with the advantage of using information from not only the next time step but from all subsequent time steps~\citep{lee2019global}. Reinforcement learning is also used for building algorithmic trading systems~\citep{deng2016deep}.
    \item \textit{Transfer learning.} Transfer learning can be used in training deep neural networks with a small amount of training data and a reduced training time, by tuning the pre-trained model on a larger training dataset, \textit{e.g.}, ~\cite{nguyen2019novel} trains a LSTM base model on 50 stocks and transfers parameters for the prediction model on KOSPI 200 or S\&P 500. 
\end{itemize}

\begin{center}
\begin{longtable}{p{0.3\textwidth}p{0.3\textwidth}p{0.4\textwidth}}
\caption{List of other types of models.} \label{tab:other} \\
\hline Article & Prediction Model & Baselines \\ \hline 
\endfirsthead

\multicolumn{3}{c}%
{{\bfseries \tablename\ \thetable{} -- continued from previous page}} \\
\hline Article & Prediction Model & Baselines \\ \hline 
\endhead

\hline \multicolumn{3}{r}{{Continued on next page}} \\ \hline
\endfoot

\hline
\endlastfoot

\hline
\multicolumn{3}{l}{Generative Adversarial Network} \\
\hline
\cite{zhou2018stock} & GAN & ARIMA-GARCH, ANN, SVM \\
\cite{zhang2019stock} & GAN & LSTM, ANN, SVM \\

\hline
\multicolumn{3}{l}{Graph Neural Network} \\
\hline
\cite{chen2018incorporating} & GCNN & LR, LSTM \\
\cite{feng2019temporal} & TGC & SFM RNN~\citep{zhang2017stock}, LSTM \\
\cite{kim2019hats} & HGAN & MLP, CNN, LSTM, GCNN~\citep{chen2018incorporating}, TGC~\citep{feng2019temporal} \\
\cite{matsunaga2019exploring} & GNN & LSTM \\

\hline
\multicolumn{3}{l}{Capsule Network} \\
\hline
\cite{liu2019transformer} & CapTE & TSLDA~\citep{nguyen2015topic}, HAN~\citep{hu2018listening}, HCAN~\citep{xu2018stock}, CH-RNN~\citep{wu2018hybrid} \\

\hline
\multicolumn{3}{l}{Reinforcement Learning} \\
\hline
\cite{lee2019global} & Deep Q-Network+CNN & FC network, CNN, and LSTM, momentum, MACD\\

\hline
\multicolumn{3}{l}{Transfer Learning} \\
\hline
\cite{hoseinzade2019u} & Transfer Learning+CNN & PCA+ANN~\citep{zhong2017forecasting}, ANN~\citep{kara2011predicting}, CNN~\citep{gunduz2017intraday}, CNN~\citep{hoseinzade2019cnnpred} \\
\cite{nguyen2019novel} & Transfer Learning+LSTM & SVM, RF, KNN, LSTM~\citep{fischer2018deep} \\

\end{longtable}
\end{center}

\begin{figure}[!htb]
    \centering
    \includegraphics[width=\textwidth]{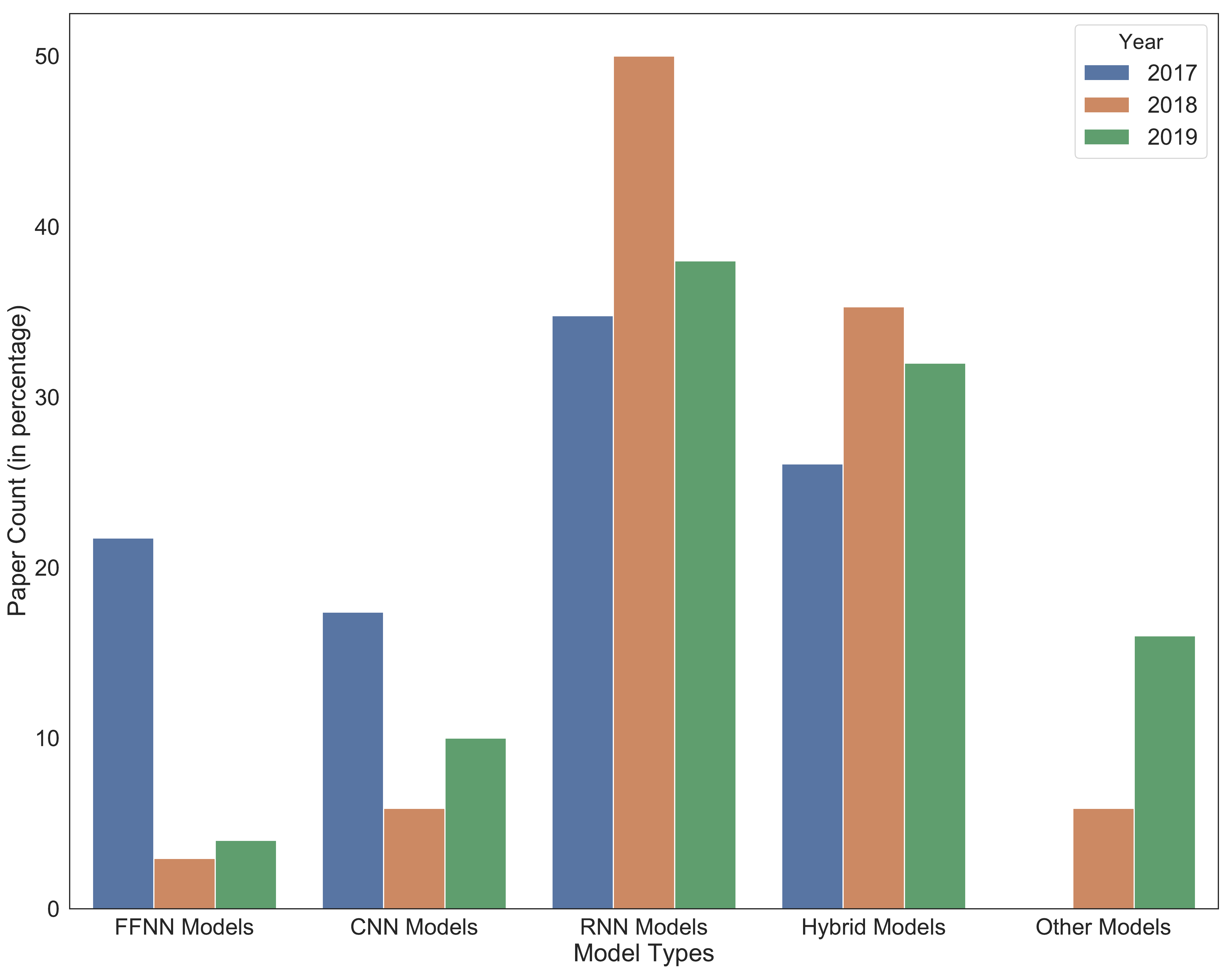}
    \caption{The usage of different models.}
    \label{fig:models}
\end{figure}

\begin{figure}[!htb]
    \centering
    \includegraphics[width=\textwidth]{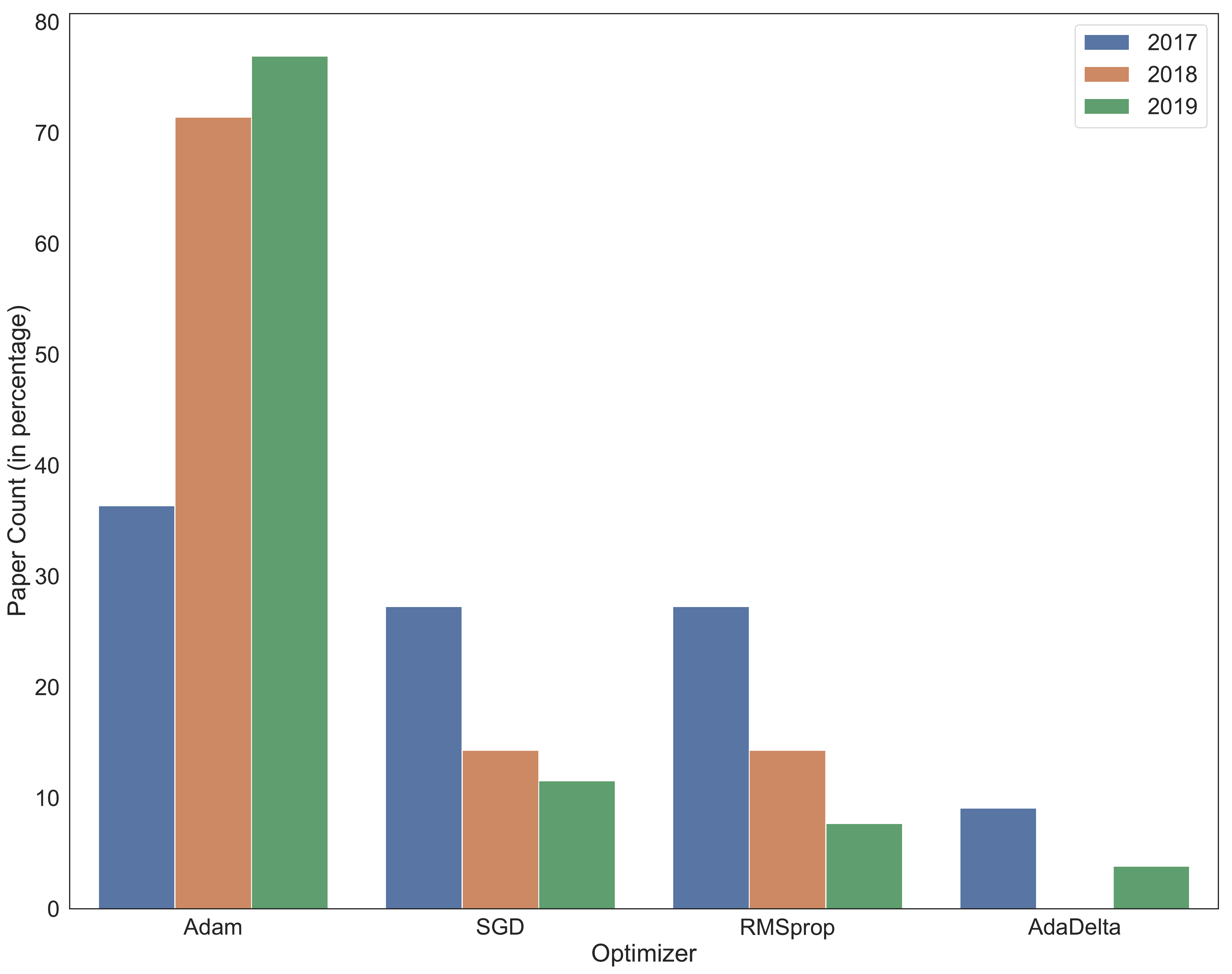}
    \caption{The usage of different optimizers.}
    \label{fig:optimizer}
\end{figure}

We show the change trend of models used in the past three years in Figure~\ref{fig:models}. RNN Models are used for the most, but the ratio drops with the emerging new models in 2019. We also show the change of common optimizers used in our surveyed papers in Figure~\ref{fig:optimizer}, which include Adam~\citep{kingma2014adam}, stochastic gradient descent (SGD), RMSprop~\citep{tieleman2012lecture}, AdaDelta~\citep{zeiler2012adadelta}. Adam has been used the most for stock prediction, which is a combination of RMSprop and stochastic gradient descent with momentum and presents several benefits, \textit{e.g.}, computationally efficiency and little memory requirement.

\begin{figure}[!htb]
    \centering
    \includegraphics[width=\textwidth]{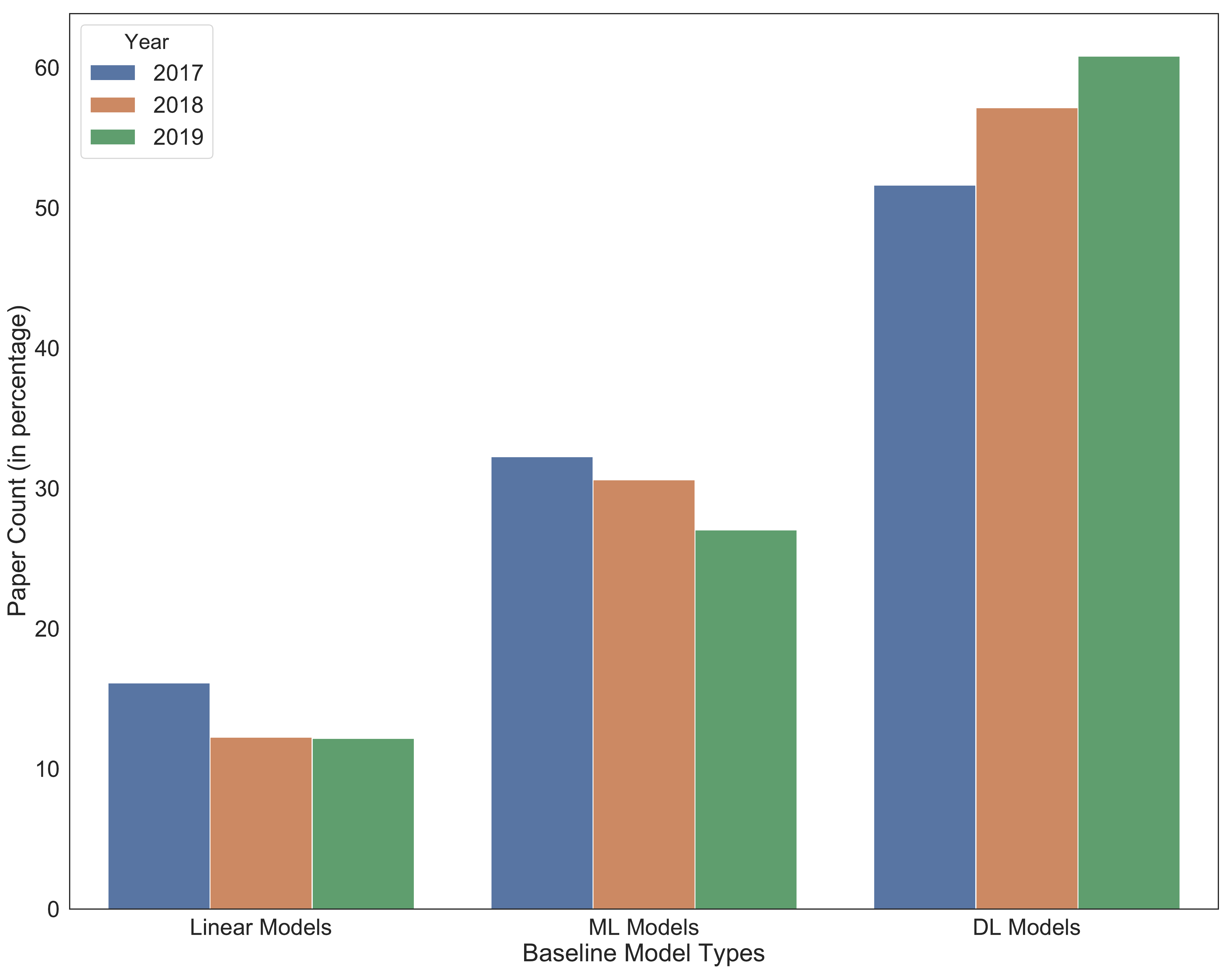}
    \caption{The usage of different baselines.}
    \label{fig:baselines}
\end{figure}

For the baselines used in the survey studies, both linear, machine learning and deep learning models are covered. The change of baseline models used is shown in Figure~\ref{fig:baselines}. With the further exploration of deep learning models for stock prediction, their ratio as baselines keeps increasing in the past three years.

\subsection{Model Evaluation}
\label{sec:evaluation}
In this part, we category the evaluation metrics for the prediction models mentioned in the last part into four types:

\begin{itemize}
    \item \textit{Classification metrics}. Classification metrics are used to measure the model's performance on movement prediction, which is modeled as a classification problem. Common used metrics include accuracy (which is the correct number of prediction for directional change), precision, recall, sensitivity, specificity, F1 score, macro-average F-score, Matthews correlation coefficient (which is a discrete case for Pearson correlation coefficient), average AUC score (area under Receiver Operating Characteristic curves~\citep{fawcett2006introduction}), Theil's U coefficient, hit ratio, average relative variance, etc. Confusion matrices and boxplots for daily accuracy are also used for classification performance analysis~\citep{guang2019multi, zhong2017forecasting, zhang2019deeplob}.
    \item \textit{Regression metrics.} Regression metrics are used to measure the model's performance on stock/index price prediction, which is modeled as a regression problem. Common used metrics include mean absolute error (MAE), root mean absolute error (RMAE), mean squared error (MSE), normalized MSE (NMSE), root mean squared error (RMSE), relative RMSE, normalized RMSE (NRMSE), mean absolute percentage error (MAPE), root mean squared relative error (RMSRE), mutual information, $R^2$ (which is the coefficient of determination).
    \item Profit Analysis. Profit analysis evaluates whether the predicted-based trading strategy can bring a profit or not. It is usually evaluated from two aspects, the return and the risk. The return is the change in value on the stock portfolio and the risk can be evaluated by maximum drawdown~\citep{zhou2019emd2fnn}, which is the largest peak-to-trough decline in the value of a portfolio and represents the max possible loss, or the annualized volatility\citep{karathanasopoulos2019forecasting}. Sharpe Ratio is a comprehensive metric with both the return and risk into consideration, which is the average return earned in excess of the risk-free per unit of volatility~\citep{sharpe1994sharpe}. More detailed analysis about the transactions is given in~\cite{sezer2018algorithmic, sezer2019financial}.
    \item Significance Analysis. In order to determine if there is significant difference in terms of predictions when comparing the deep learning models to the baselines, Kruskal-Wallis~\citep{kruskal1952use} and Diebold-Mariano~\citep{diebold2002comparing} tests can be used to test the statistical significance, which decides a statistically better model. They are not used often for stock prediction, with only a few studies in 2019~\citep{kumar2019predicting, zhang2019deeplob}.
\end{itemize}

The detailed list of studies using each metrics (as well as the metrics' abbreviations) is shown in Table~\ref{tab:metrics}.

\begin{center}
\begin{longtable}{p{0.4\textwidth}p{0.6\textwidth}}
\caption{Article Lists archived by different evaluation metrics.}
\label{tab:metrics} \\
\hline Metrics & Article List \\ \hline 
\endfirsthead

\multicolumn{2}{c}%
{{\bfseries \tablename\ \thetable{} -- continued from previous page}} \\
\hline Metrics & Article List \\ \hline 
\endhead

\hline \multicolumn{2}{r}{{Continued on next page}} \\ \hline
\endfoot

\hline
\endlastfoot

\hline
\multicolumn{2}{l}{Classification Metrics} \\
\hline
Accuracy & \cite{de2013applying, niaki2013forecasting, ding2014using, ding2015deep, chen2017double,dingli2017financial, huynh2017new, li2017combining, liu2017stock, nelson2017stock, selvin2017stock, singh2017stock, sun2017stacked, vargas2017deep, weng2017stock, yang2017stock, zhang2017data, zhao2017time, assis2018restricted, chen2018artificial, chen2018incorporating, chen2018leveraging, cheng2018applied, fischer2018deep, gao2018improving, hu2018listening, huang2018tensor, liu2018hierarchical, liu2018numerical, matsubara2018stock, minh2018deep, oncharoen2018deep, pang2018innovative, sezer2018algorithmic, tang2018stock, tran2018temporal, wu2018hybrid, yang2018multi, zhou2018stock, araujo2019deep, chen2019exploring, deng2019knowledge, feng2019enhancing, guang2019multi, karathanasopoulos2019forecasting, lee2019global, li2019dp, li2019multitask, liu2019combining, liu2019transformer, long2019deep, merello2019ensemble,nguyen2019novel, sanboon2019deep, song2019study, sun2019exploiting, tan2019tensor, tang2019learning, wang2019ean} \\
Precision & \cite{gunduz2017intraday, li2017sentiment, nelson2017stock, tsantekidis2017forecasting, tsantekidis2017using, cheng2018applied, minh2018deep, tran2018temporal, sezer2018algorithmic, li2019multitask, wang2019clvsa, zhang2019deeplob} \\
Recall & \cite{gunduz2017intraday, li2017sentiment, nelson2017stock, tsantekidis2017forecasting, tsantekidis2017using, cheng2018applied, minh2018deep, tran2018temporal, sezer2018algorithmic, li2019multitask, zhang2019deeplob} \\
Sensitivity & \cite{sim2019deep} \\
Specificity & \cite{sim2019deep} \\
F1 score (F1) & \cite{gunduz2017intraday, li2017sentiment, nelson2017stock, sun2017stacked, tsantekidis2017forecasting, tsantekidis2017using, zhang2017data, cheng2018applied, jiang2018cross, sezer2018algorithmic, tran2018temporal, deng2019knowledge, guang2019multi, li2019multitask, wang2019ean, zhang2019deeplob} \\
Macro-average F-score (MAFS) & \cite{hoseinzade2019cnnpred, hoseinzade2019u} \\
Matthews Correlation Coefficient (MCC) & \cite{ding2014using, ding2015deep, singh2017stock, huang2018tensor, matsubara2018stock, xie2018recurrent, cao2019stock, feng2019enhancing, liu2019transformer, merello2019ensemble, tan2019tensor, tang2019learning} \\
Average AUC Score (AUC) & \cite{ballings2015evaluating, jiang2018cross, borovkova2019ensemble, zhang2019deeplob} \\
Theil's U Coefficient (Theil's U) & \cite{de2013applying, bao2017deep, yan2018financial, araujo2019deep} \\
Hit Ratio (Hit) & \cite{singh2017stock, hu2018predicting, sim2019deep} \\
Average Relative Variance (ARV) & \cite{araujo2019deep} \\

\hline
\multicolumn{2}{l}{Regression Metrics} \\
\hline

Mean Absolute Error (MAE) & \cite{patel2015predicting, chong2017deep, li2017time, qin2017dual, althelaya2018evaluation, althelaya2018stock, baek2018modaugnet, chen2018leveraging, chung2018genetic, gao2018improving, hossain2018hybrid, lei2018wavelet, al2019forecasting, cao2019financial, chen2019hybrid, ding2019study, jin2019stock, karathanasopoulos2019forecasting, liu2019non, nikoustock, nguyen2019predicting, tang2019learning, yu2019ceam, zhang2019stock, zhou2019emd2fnn} \\
Root Mean Absolute Error (RMAE) & \cite{kim2019forecasting} \\
Mean Squared Error (MSE) & \cite{patel2015predicting, li2017time, zhang2017stock, zhong2017forecasting, baek2018modaugnet, chung2018genetic, hollis2018comparison, hossain2018hybrid, liu2018numerical, pang2018innovative, zhan2018stock, araujo2019deep, eapen2019novel, feng2019temporal, karathanasopoulos2019forecasting, nguyen2019predicting, nikoustock, sachdeva2019effective, stoean2019deep} \\
Normalized MSE (NMSE) & \cite{chong2017deep} \\
Root Mean Squared Error (RMSE) & \cite{de2013applying, chong2017deep, lee2017predict, li2017time, qin2017dual, singh2017stock, althelaya2018evaluation, althelaya2018stock, chen2018artificial, chen2018leveraging, gao2018improving, lei2018wavelet, siami2018comparison, al2019forecasting, cao2019financial, cao2019stock, chen2019a, chen2019hybrid, jin2019stock, karathanasopoulos2019forecasting, kim2019forecasting, liu2019non, nikoustock, sethia2019application, siami2019comparative, sun2019exploiting, tang2019learning, yu2019ceam, zhang2019stock, zhou2019emd2fnn} \\
Relative RMSE (rRMSE) & \cite{patel2015predicting, nguyen2019predicting} \\
Normalized RMSE (NRMSE) & \cite{kumar2019predicting} \\
Mean Absolute Percentage Error (MAPE) & \cite{de2013applying, ticknor2013bayesian, patel2015predicting, bao2017deep, li2017combining, qin2017dual, singh2017stock, yang2017stock, baek2018modaugnet, chen2018artificial, chen2018leveraging, chung2018genetic, gao2018improving, hossain2018hybrid, lei2018wavelet, wu2018adaboost, xie2018recurrent, yan2018financial, araujo2019deep, cao2019financial, chen2019hybrid, jin2019stock, karathanasopoulos2019forecasting, kim2019forecasting, kumar2019predicting, mohan2019stock, nguyen2019predicting, sachdeva2019effective, yu2019ceam, zhang2019stock, zhou2019emd2fnn} \\
Root Mean Squared Relative Error (RMSRE) & \cite{zhou2018stock} \\
Mutual Information (MUL) & \cite{chong2017deep} \\
$R^2$ & \cite{bao2017deep, althelaya2018evaluation, althelaya2018stock, al2019forecasting, chen2019hybrid, jin2019stock, liu2019non, sethia2019application} \\

\hline
\multicolumn{2}{l}{Profit Metrics} \\
\hline

Return & \cite{niaki2013forecasting, ding2015deep, bao2017deep, chen2017double, lee2017predict, singh2017stock, zhong2017forecasting, fischer2018deep, hu2018listening, matsubara2018stock, oncharoen2018deep,sezer2018algorithmic, wu2018hybrid, xie2018recurrent, yang2018multi, chen2019exploring, feng2019temporal, hoseinzade2019cnnpred, karathanasopoulos2019forecasting, kim2019forecasting, lee2019global, long2019deep, matsunaga2019exploring, merello2019ensemble, sezer2019financial, stoean2019deep, song2019study, sun2019exploiting, wang2019clvsa, zhang2019stock, zhou2019emd2fnn} \\
Maximum Drawdown & \cite{zhou2019emd2fnn} \\
Annualized Volatility & \cite{karathanasopoulos2019forecasting} \\
Sharpe Ratio & \cite{chen2017double, fischer2018deep, sezer2018algorithmic, hoseinzade2019cnnpred, karathanasopoulos2019forecasting, matsunaga2019exploring, merello2019ensemble, stoean2019deep, wang2019clvsa, zhou2019emd2fnn} \\

\hline
\multicolumn{2}{l}{Significance Analysis} \\
\hline

Kruskal-Wallis Test & \cite{zhang2019deeplob} \\
Diebold-Mariano Test & \cite{kumar2019predicting} \\

\end{longtable}
\end{center}

\begin{figure}[!htb]
    \centering
    \includegraphics[width=\textwidth]{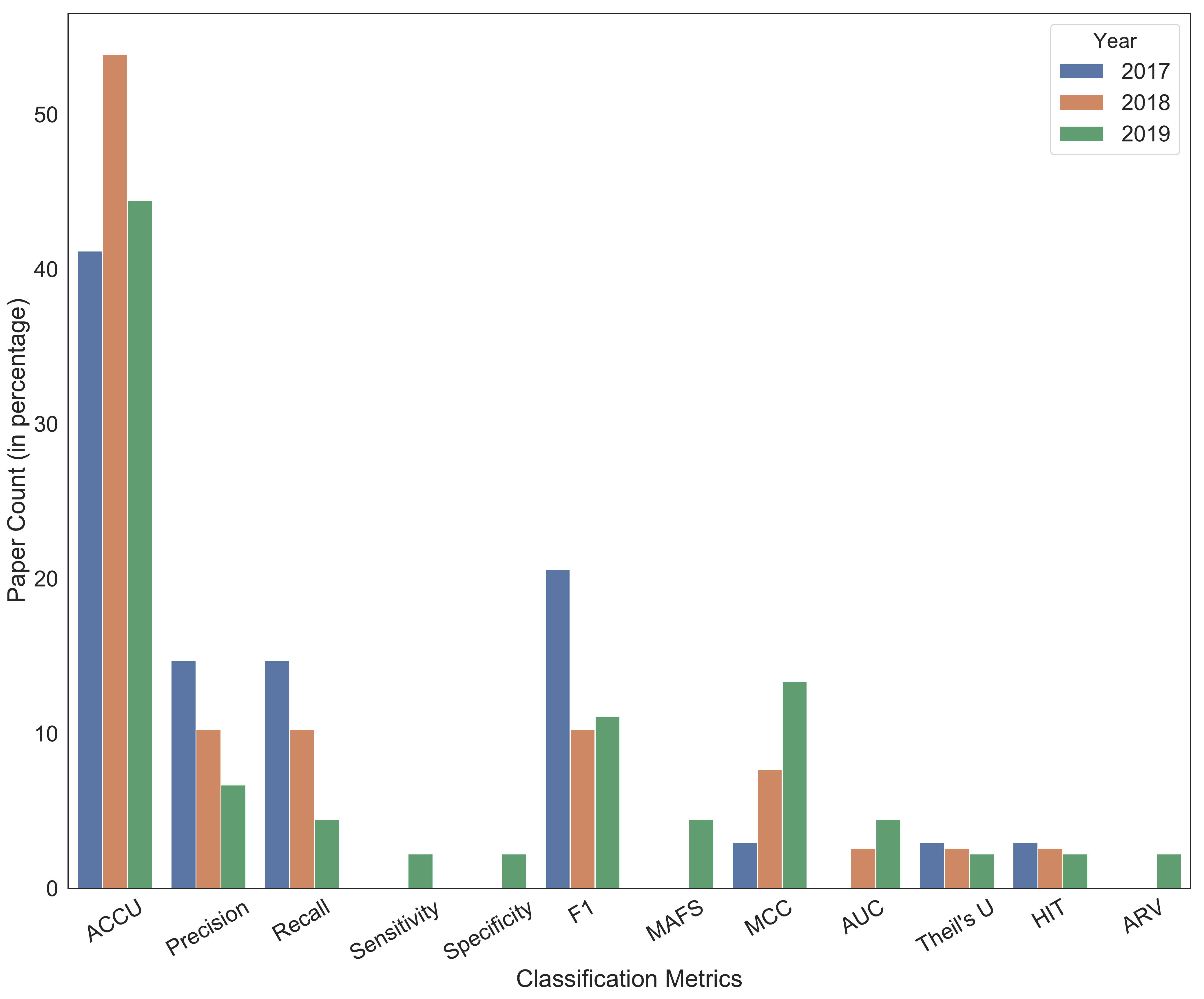}
    \caption{The usage of different classification metrics.}
    \label{fig:metrics_classification}
\end{figure}

\begin{figure}[!htb]
    \centering
    \includegraphics[width=\textwidth]{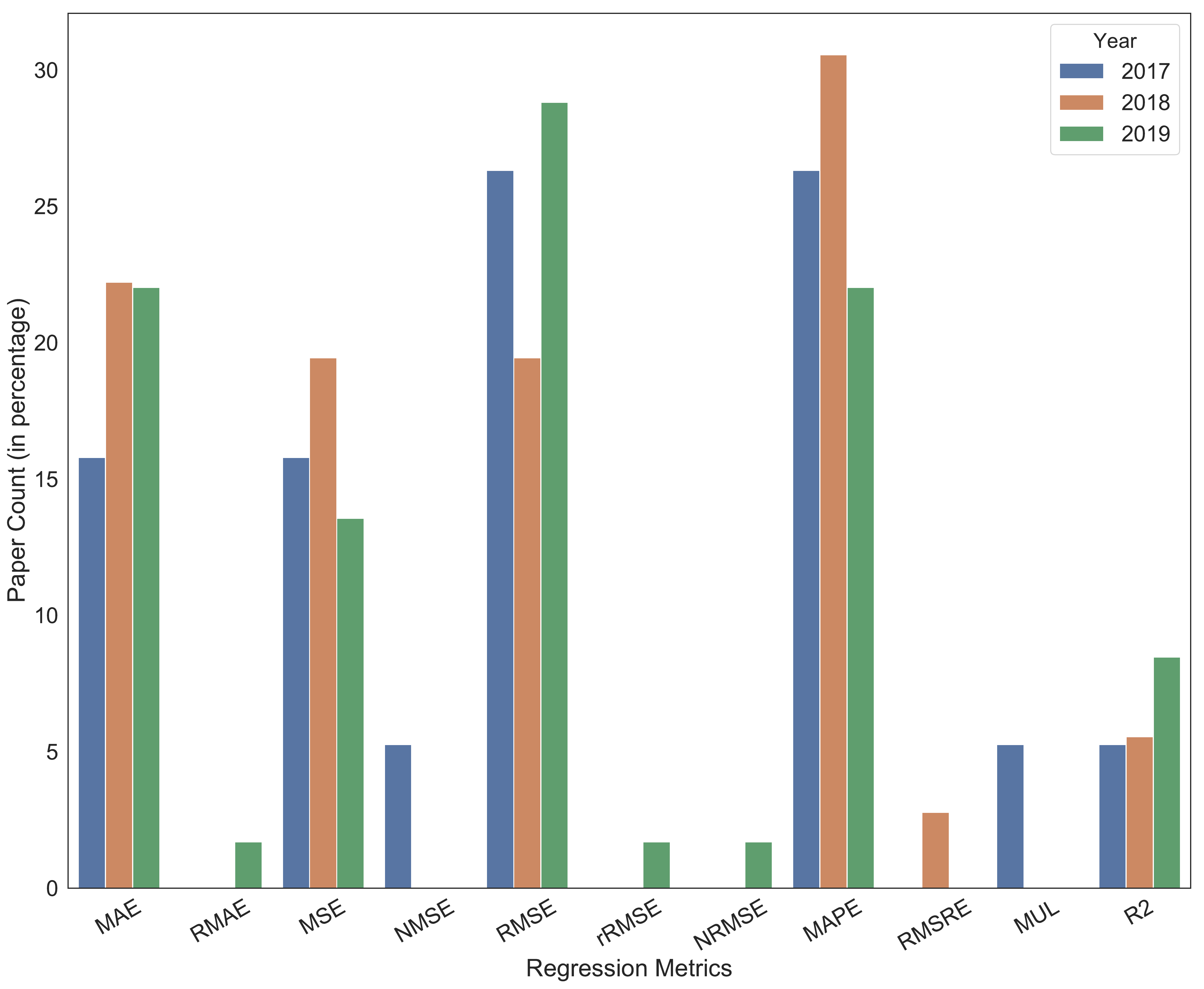}
    \caption{The usage of different regression metrics.}
    \label{fig:metrics_regression}
\end{figure}

We further show the change of classification and regression metrics in Figure~\ref{fig:metrics_classification} and Figure~\ref{fig:metrics_regression}. For classification metrics, accuracy and F1 score are the most often used, followed by precision, recall, and MCC. For regression metrics, RMSE and MAPE are the most often used, followed by MAE and MSE.

\section{Implementation and Reproducibility}
    \label{sec:implementation}
    
\subsection{Implementation}
In this section, we pay a special attention to the implementation details of the papers we survey, which is less discussed before in previous surveys.

\begin{figure}[!htb]
    \centering
    \includegraphics[width=\textwidth]{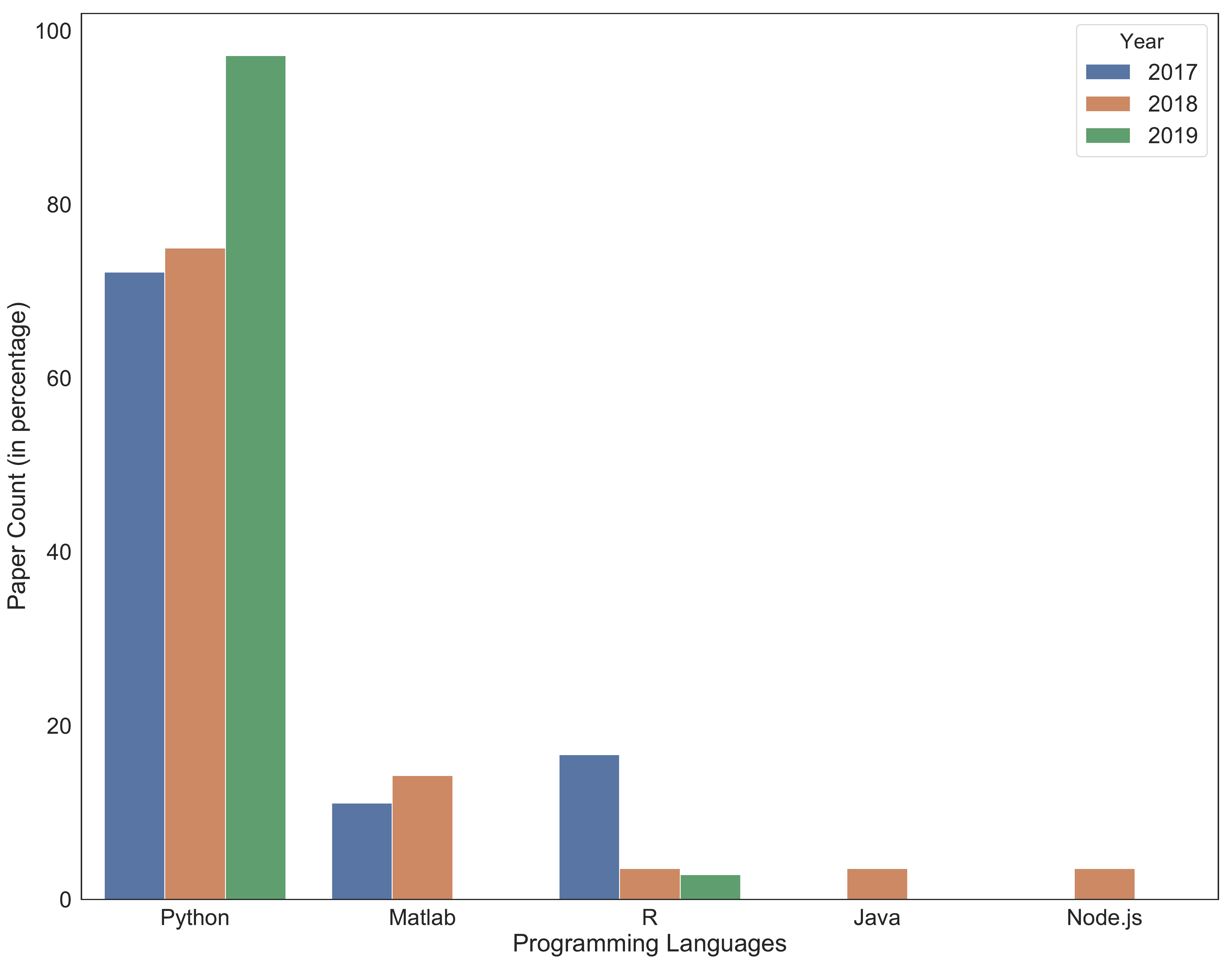}
    \caption{The trend of different programming languages for stock market prediction.}
    \label{fig:language}
\end{figure}

We firstly investigate the programming language used for the implementation of machine learning and deep learning models. Among them, Python is becoming the dominant choice in the past three years as shown in Figure~\ref{fig:language}, which provides a bunch of packages and frameworks for model implementation purpose, e.g., Keras~\footnote{\url{http://keras.io/}. Keras has been incorporated in TensorFlow 2.0 and higher versions.}, TensorFlow~\footnote{\url{https://www.tensorflow.org/}}, PyTorch~\footnote{\url{https://pytorch.org/}}, Theano~\footnote{\url{http://deeplearning.net/software/theano/}. It is a discontinued project and not recommended for further use.}, scikit-learn~\footnote{\url{https://scikit-learn.org/stable/index.html}}. Other choices include R, Matlab, Java, etc. We show the specific paper list using different programming languages and frameworks in Table~\ref{tab:language}. From Table~\ref{tab:language}, Keras and TensorFlow are the dominant frameworks for deep learning-based stock market prediction research. For further reference, the readers may refer to~\cite{hatcher2018survey} for a comprehensive introduction of deep learning tools.

\begin{center}
\begin{longtable}{llp{0.6\textwidth}}
\caption{List of programming languages and frameworks for stock market prediction.} \label{tab:language} \\
\hline Language & Framework & Articles \\ \hline 
\endfirsthead

\multicolumn{3}{c}%
{{\bfseries \tablename\ \thetable{} -- continued from previous page}} \\
\hline Language & Framework & Articles \\ \hline 
\endhead

\hline \multicolumn{3}{r}{{Continued on next page}} \\ \hline
\endfoot

\hline
\endlastfoot

Python & Keras & \cite{zhang2017stock, sezer2018algorithmic, hossain2018hybrid, eapen2019novel, siami2018comparison, fischer2018deep, althelaya2018evaluation, baek2018modaugnet, althelaya2018stock, minh2018deep, sezer2019financial, hoseinzade2019cnnpred, al2019forecasting, liu2019non, nguyen2019novel, chen2019exploring, sachdeva2019effective, lee2017predict, li2017sentiment, nguyen2019predicting, gunduz2017intraday, huynh2017new, chen2019hybrid, zhang2019deeplob} \\
Python & TensorFlow & \cite{feng2019temporal, kim2019hats, sezer2018algorithmic, dingli2017financial, sim2019deep, song2019study, liu2019anticipating, xu2018stock, eapen2019novel, zhan2018stock, huang2018tensor, fischer2018deep, althelaya2018evaluation, gao2018improving, althelaya2018stock, zhao2017time, vargas2017deep, lee2019global, sezer2019financial, al2019forecasting, qin2017dual, liu2019non, chen2019exploring, nguyen2019predicting, ding2019study, zhang2019deeplob, borovkova2019ensemble} \\
Python & PyTorch & \cite{yu2019ceam} \\
Python & Theano & \cite{zhang2017stock, chong2017deep, siami2018comparison, pang2018innovative, tsantekidis2017forecasting} \\
Python & scikit-learn & \cite{liu2019anticipating, eapen2019novel, gao2018improving, zhao2017time, nguyen2019novel} \\
Python & N/A & \cite{cao2019financial, wu2018hybrid, li2019multitask, siami2019comparative, jin2019stock, chen2018leveraging} \\
Matlab & N/A & \cite{ticknor2013bayesian, zhong2017forecasting, singh2017stock, wu2018adaboost, lei2018wavelet, pang2018innovative, chen2018artificial} \\
R & nnet & \cite{ballings2015evaluating, chen2017double} \\
R & N/A & \cite{weng2017stock, singh2017stock, fischer2018deep, kumar2019predicting} \\
Java & N/A & \cite{pang2018innovative} \\
Node.js & N/A & \cite{assis2018restricted} \\

\end{longtable}
\end{center}

Deep learning models require a larger amount of computation for training, and GPU has been used to accelerate the convolutional operations involved. With the need of processing multiple types of input data, especially the text data, the need for GPU would keep increasing in this research area. We give a list of different types of GPU used in the surveyed papers in Table~\ref{tab:gpu}. Cloud computing is another solution when GPU is not available locally. There are many commercial choices of cloud computing services, e.g., Amazon Web Services~\footnote{\url{https://aws.amazon.com/}}, Google Cloud~\footnote{\url{https://cloud.google.com/}}, and Microsoft Azure~\footnote{\url{https://azure.microsoft.com/}}. However, they are not widely adopted in current study of stock market prediction and no previous study covered in this survey mentions the usage of cloud service explicitly.

\begin{table}[!htb]
    \centering
    \caption{List of GPUs used for stock market prediction.}
    \label{tab:gpu}
    \begin{tabular}{lp{0.6\textwidth}}
        \hline
        GPU Type (all from NVIDIA) & Articles \\
        \hline
        Tesla P100 & \cite{zhang2019deeplob} \\
        GeForce GTX 1060 & \cite{song2019study} \\
        GeForce GTX 1080 Ti & \cite{eapen2019novel} \\
        Quadro P2000 & \cite{nguyen2019predicting} \\
        TITAN X & \cite{chong2017deep, minh2018deep, chen2019exploring} \\
        TITAN Xp & \cite{chen2019exploring} \\
        TITAN RTX & \cite{wang2019clvsa} \\
        Not specified & \cite{xu2018stock, fischer2018deep, guang2019multi} \\
        \hline
    \end{tabular}
\end{table}

\subsection{Result Reproducibility}
While deep learning techniques have been proved to be effective in many different problems and most of the previous studies have proven their effectiveness for time series forecasting problems, while there are still doubts and concerns, \textit{e.g.}, in the M4 open forecasting competition with 100,000 time series which started on Jan 1, 2018 and ended on May 31, 2018, statistical approaches outperform pure ML methods~\citep{makridakis2018m4} and there are similar results with the dataset of the earlier M3 competition~\citep{makridakis2018statistical}. These studies also question the reproducibility and replicability in the previous papers which use ML methods.

While it is beyond the scope of this study to check the result of each paper, we instead investigate the data and code availability of the surveyed papers, which are two important aspects for the result reproducibility. Some of the source journals would require or recommend the data and code submitted as supplementary files for peer review, e.g., PLOS ONE. In other cases, the authors would share their data and code proactively, for the consideration that following works can easily use them as baselines, which gains a higher impact for their publications.

\subsubsection{Data Availability}
There are many free data sources on the Internet for the research purpose of stock market prediction. For historical price and volume, the first choice should be the widely used Yahoo! Finance~\footnote{\url{https://finance.yahoo.com/}}, which provides free access to data including stock quotes, up-to-date news, international market data, etc., and has been mentioned at least in 25 out of 124 papers. Other similar options include Tushare~\footnote{\url{https://tushare.pro/}}, which can be used to crawl historical data of China stocks. Some stock markets would also provide the download service of historical data on their official websites. For macroeconomic indicators, International Monetary Fund (IMF)~\footnote{\url{https://www.imf.org/}} and World Bank~\footnote{\url{https://www.worldbank.org/}} are good choices to explore. For financial news, previous studies would crawl some major news sources, e.g., CNBC~\footnote{\url{https://www.cnbc.com/}}, Reuters~\footnote{\url{https://www.reuters.com/}}, Wall Street Journal~\footnote{\url{https://www.wsj.com/}}, Fortune~\footnote{\url{https://fortune.com/}}, etc. Social networking websites, e.g., Twitter~\footnote{\url{https://twitter.com/}} and Sina Weibo~\footnote{\url{https://www.weibo.com/}}, provide web Application Programming Interface (API) for the access of their data (usually preprocessed and anonymous). And researchers could filter the financial related tweet using companies' names as keywords. For relational data, Wikidata~\footnote{\url{https://www.wikidata.org/wiki/Wikidata:Main_Page}} provides relations between companies such as supplier-consumer relation and ownership relation.

There are many commercial choices too, e.g., Bloomberg~\footnote{\url{https://www.bloomberg.com/}}, Wind~\footnote{\url{https://www.wind.com.cn/}}, Quantopian~\footnote{\url{https://www.quantopian.com/}}, Investing.com~\footnote{\url{https://www.investing.com/}}. Online brokers such as Interactive Brokers~\footnote{\url{https://www.interactivebrokers.com/}} also provide data-related services. There are also some well-established databases for research purpose and contain many different types of financial data, which can be used for stock market prediction and other financial problems. One example is the CSMAR database~\footnote{\url{http://us.gtadata.com/}}, which provides financial statements and stock trading data for Chinese companies, including balance sheets, income statements, cash flows, stock prices and returns, market returns and indices, and other data on Chinese equities. Another example is Wharton Research Data Services (WRDS)~\footnote{\url{https://wrds-web.wharton.upenn.edu/}}, which provides access to many financial, accounting, banking, economics, marketing, and public policy databases through a uniform, web-based interface.

Data competition websites, e.g., Kaggle~\footnote{\url{https://www.kaggle.com/}}, are also becoming a good choice of data repository for stock market prediction. And quantitative companies could collaborate with these websites to host stock market prediction competitions, e.g., Two Sigma Financial Modeling Challenge~\footnote{\url{https://www.kaggle.com/c/two-sigma-financial-modeling}}, which is organized by a hedge fund named Two Sigma~\footnote{\url{https://www.twosigma.com/}}.

Even though most of the data sources are available on the Internet, it would be more convenient for replicability if the authors could release the exact dataset they use. In Table~\ref{tab:data}, we list those with the data description and link, for those data which is hosted in software host websites such as Github~\footnote{\url{https://github.com/}}, cloud services, researcher's own website, and data competition websites such as Kaggle. For the mid-price prediction of limit order book data, there is a benchmark dataset provided by~\cite{ntakaris2017benchmark} and has been used in the following studies~\citep{tran2018temporal}.

\begin{center}
\begin{longtable}{lp{0.6\textwidth}}
\caption{List of articles with public available data links.} \label{tab:data} \\
\hline Articles & Data Description \& Link \\ \hline 
\endfirsthead

\multicolumn{2}{c}%
{{\bfseries \tablename\ \thetable{} -- continued from previous page}} \\
\hline Articles & Data Description \& Link \\ \hline 
\endhead

\hline \multicolumn{2}{r}{{Continued on next page}} \\ \hline
\endfoot

\hline
\endlastfoot
\cite{bao2017deep} & CSI 300, Nifty 50, Hang Seng index, Nikkei 225, S\&P500 and DJIA index from Jul-01-2008 to Sep-30-2016. Link: \url{https://doi.org/10.6084/m9.figshare.5028110} \\
\cite{qin2017dual} & One-minute stock prices of 104 corporations under NASDAQ 100 and the index value of NASDAQ 100 from Jul-26-2016 to Apr-28-2017. Link: \url{http://cseweb.ucsd.edu/~yaq007/NASDAQ100_stock_data.html} \\
\cite{zhang2017stock} & The daily opening prices of 50 stocks in US among 10 sectors from 2007 to 2016. Link: \url{https://github.com/z331565360/State-Frequency-Memory-stock-prediction/tree/master/dataset} (also used in~\cite{feng2019enhancing})\\
\cite{hollis2018comparison} & Historical data and Thomson Reuters news since 2007. Link: \url{https://www.kaggle.com/c/two-sigma-financial-news} \\
\cite{huang2018tensor} & 78 A-share stocks in CSI 100 and 13 popular HK stocks in the year 2015 and 2016. Financial web news dataset: \url{https://pan.baidu.com/s/1mhCLJJi}; Guba dataset: \url{https://pan.baidu.com/s/1i5zAWh3} \\
\cite{wu2018hybrid} & Prices and Twitter for 47 stocks listed in S\&P 500 from January 2017 to November 2017. Link: \url{https://github.com/wuhuizhe/CHRNN} (also used in~\cite{liu2019transformer}) \\
\cite{xu2018stock} & Historical data of 88 high-trade-volume-stocks in NASDAQ and NYSE markets from Jan-01-2014 to Jan-01-2016. Link: \url{https://github.com/yumoxu/stocknet-dataset} (also used in ~\cite{feng2019enhancing}) \\
\cite{feng2019enhancing} & Data from previous studies~\citep{zhang2017stock, xu2018stock}. Link: \url{https://github.com/fulifeng/Adv-ALSTM/tree/master/data} \\
\cite{feng2019temporal} & Historical price data, Sector-industry relations, and Wiki relations between their companies such as supplier-consumer relation and ownership relation for 1,026 NASDAQ and 1,737 NYSE stocks from Jan-03-2017 to Dec-08-2017. Link: \url{https://github.com/fulifeng/Temporal_Relational_Stock_Ranking/tree/master/data} \\
\cite{liu2019non} & 6 top banks in US from 2008 to 2016. Link: \url{https://www.kaggle.com/rohan8594/stock-data} \\
\cite{kim2019forecasting} & Minute SPY ticker data from Oct-14-2016 to Oct-16-2017. Link: \url{https://dx.doi.org/10.6084/m9.figshare.7471568} \\
\cite{sim2019deep} & Minute data of the S\&P 500 index from 10:30 pm on Apr-03-2017, to 2:15 pm on May-02-2017. Link: \url{https://www.kesci.com/home/dataset/5bbdc2513631bc00109c29a4/files} \\
\cite{stoean2019deep} & 25 companies listed under the Romanian stock market from Oct-16-1997 to Mar-13-2019. Link: \url{https://doi.org/10.6084/m9.figshare.7976144.v1} \\
\cite{liu2019anticipating} & News headlines from Thomson Reuters and Cable News Network for 6 stocks in US markets. Link: \url{https://github.com/linechany/knowledge-graph} \\
\end{longtable}
\end{center}

\subsubsection{Code Availability}
Github has been the mainstream platform of hosting source code in the computer science field. However, only a small number of studies would release their code for now, in the area of stock market prediction. In Table~\ref{tab:code}, we list the articles with public code repositories. A short description of each method is mentioned, and the details can be found in Section~\ref{sec:workflow} and the original documents.

\begin{table}[!htb]
    \centering
    \caption{List of articles with public code links.}
    \label{tab:code}
    \begin{tabular}{lp{0.6\textwidth}}
        \hline
        Articles & Method Description \& Link \\
        \hline
        \cite{weng2017stock} & Artificial neural networks (ANN), decision trees (DT), and support vector machines (SVM) in R. Link: \url{https://github.com/binweng/ShinyStock} \\
        \cite{zhang2017stock} & State frequency memory (SFM) recurrent network. Link: \url{https://github.com/z331565360/State-Frequency-Memory-stock-prediction} \\
        \cite{hu2018listening} & Hybrid attention network (HAN). Link: \url{https://github.com/gkeng/Listening-to-Chaotic-Whishpers--Code} \\
        \cite{xu2018stock} & A deep generative model named StockNet. Link: \url{https://github.com/yumoxu/stocknet-code} \\
        \cite{feng2019enhancing} & Adversarial attentive LSTM. Link: \url{https://github.com/fulifeng/Adv-ALSTM} \\
        \cite{feng2019temporal} & Relational stock ranking (RSR). Link: \url{https://github.com/fulifeng/Temporal_Relational_Stock_Ranking} \\
        \cite{kim2019hats} & Hierarchical graph attention network (HATS). Link: \url{https://github.com/dmis-lab/hats} \\
        \cite{lee2019global} & Deep Q-Network. Link: \url{https://github.com/lee-jinho/DQN-global-stock-market-prediction/} \\
        \hline
    \end{tabular}
\end{table}

\section{Future Directions}
    \label{sec:future}

Based on our review of recent works, we give some future directions in this section, which aims to bring new insight to interested researchers.

\subsection{New Models}
Different structures of neural networks are not fully studied for stock prediction, especially those who only appear in recent years. There are two steps where deep learning models involve in stock prediction, namely, Data Processing and Prediction Model in Section~\ref{sec:workflow}. While we already covered some latest effort of applying new models in this survey, e.g., the attention mechanism and generative adversarial networks, there are still a huge space to explore for new models. For example, for sentiment analysis of text data, Transformer~\citep{vaswani2017attention} and pre-trained BERT (Bidirectional Encoder Representations from Transformers)~\citep{devlin2018bert} are widely used in natural language processing, but is less discussed for financial news analysis.

\subsection{Multiple Data Sources}
Observed from our discussion in Section~\ref{sec:workflow}, it is not wise to design a stock prediction solution based on a single data source, e.g., market data, as it has been heavily used in previous studies and it would be very challenging to outperform existing solutions. A better idea is to collect and use multiple data sources, especially those which are less explored in the literature~\citep{zhou2019forecasting}.

\subsection{Cross-market Analysis}
Most of the existing studies focus on only one stock market, in the sense that stock markets differ from each other because of the trading rules, while different markets may share some common phenomenon that can be leveraged for prediction by approaches such as transfer learning. There are already a few studies showing positive results for cross-market analysis~\citep{hoseinzade2019cnnpred, lee2019global, merello2019ensemble, nguyen2019novel, hoseinzade2019u}, it is worth exploring in the following studies. In~\cite{lee2019global}, the model is trained only on US stock market data and tested on the stock market data of 31 different countries over 12 years. Even though the authors do not use the terminology of transfer learning, it is a practice of model transfer.

\subsection{Algorithmic Trading}
The prediction is not the end of the journey. Good prediction is one factor to make money in the stock market, but not the whole story. Some of the studies have evaluated the profit and risk of the trading strategies based on the prediction result, as we discussed in Section~\ref{sec:evaluation}. However, these strategies are simple and intuitive, which may be impractical limited by the trading rules. The transaction cost is often omitted or simplified, which makes the conclusion less persuasive. Another problem is the adaption for different market styles, as the training of deep learning models is time-consuming. These studies are not sufficient for building a practical algorithmic trading system. One possible direction is deep reinforcement learning, which has recent successes in a variety of applications and is also been used in a few studies for stock prediction and trading~\citep{xiong2018practical, lee2019global}. It has advantages of simulating more possible cases and making a faster and better trading choice than human traders.

\section{Conclusion}
    \label{sec:conclusion}

Inspired by the rapid development and increasing usage of deep learning models for stock market prediction, we give a review of recent progress by surveying more than 100 related published articles in the past three years. We cover each step from raw data collection and data processing to prediction model and model evaluation and present the research trend from 2017 to 2019. We also pay a special attention to the implementation of deep learning models and the reproducibility of published articles, with the hope to accelerate the process of adopting published models as baselines (maybe with new data input). With some future directions pointed up, the insight and summary in the survey would help to boost the future research in related topics. 

\section*{Acknowledgement}
    \label{sec:acknowledgement}
Weiwei Jiang: Conceptualization; Data curation; Formal analysis; Funding acquisition; Investigation; Methodology; Project administration; Resources; Software; Supervision; Validation; Visualization; Roles/Writing - original draft; Writing - review \& editing.

\bibliography{elsarticle-template}
\nocite{*}

\end{document}